\documentclass[aps,showpacs,nofootinbib,superscriptaddress]{revtex4}

\usepackage{graphicx}
\usepackage{bm}
\usepackage{amsmath}
\usepackage{amssymb}
\usepackage{hyperref}

\newcommand{\qslash}{\kern 0.2 em n\kern -0.50em /}
\newcommand{\nslash}{\kern 0.2 em n\kern -0.50em /}
\newcommand{\kslash}{\kern 0.2 em k\kern -0.45em /}
\newcommand{\lslash}{\kern 0.2 em l\kern -0.50em /}
\newcommand{\pslash}{\kern 0.2 em p\kern -0.50em /}
\newcommand{\Sslash}{\kern 0.2 em S\kern -0.50em /}
\newcommand{\Pslash}{\kern 0.2 em P\kern -0.50em /}
\newcommand{\Dslash}{\kern 0.2 em D\kern -0.65em /\kern 0.15em}

\newcommand{\bp}{\boldsymbol{p}_T}

\newcommand{\ssh}{\!\!\!/}
\newcommand{\Tr}{\operatorname*{Tr}\nolimits}

\usepackage{color}

\begin{document}

\title{Single Spin Asymmetry in transverse polarized proton production and lambda production in Semi-Inclusive DIS at twist-3}

\author{Yongliang Yang}\affiliation{School of Physics, Southeast University, Nanjing
211189, China}
\author{Wenjuan Mao}\affiliation{School of Physics and Telecommunication Engineering,
Zhoukou Normal University,
Zhoukou 466000, Henan, China}
\author{Zhun Lu}\email{zhunlu@seu.edu.cn}\affiliation{School of Physics, Southeast University, Nanjing 211189, China}

\begin{abstract}
  We study the single spin asymmetry in the transversely polarized proton production and the lambda hyperon production in semi-inclusive deep inelastic scattering with a $\sin \phi_{S_h}$ modulation, in which $S_h$ is the azimuthal angle of the transverse spin of the final hadron. The theoretical interpretation for the asymmetry is presented as the convolution of the twist-3 quark transverse momentum dependent distributions and twist-2 fragmentation functions. We investigate the role of the twist-3 distribution functions $h(x,\bm p_T^2)$, $f^\perp(x,\bm p_T^2)$ and $g^\perp(x,\bm p_T^2)$ and the twist-2 fragmentation functions $H_1(x,\bm k_T^2)$, $D_{1T}^\perp(x,\bm k_T^2)$ and $G_{1T}(x,\bm k_T^2)$ in this asymmetry. Using spectator model, we compute these distribution and fragmentation functions and estimate the $\sin \phi_{S_h}$ asymmetry for the proton production as well as for the lambda production at JLab with a 12 $\mathrm{GeV}$ electron beam. The asymmetry at the COMPASS kinematics is also predicted.
  By comparing different sources for the asymmetry, we conclude that it is feasible to access the distribution function $h$ and the fragmentation function $H_1$ through measuring the asymmetry in the proton production, whereas the $f^\perp \,D_{1T}^\perp$ term  might be probed in the lambda production at JLab 12 GeV.
\end{abstract}


\maketitle

\section{introduction}

The single spin asymmetry (SSA) is a powerful tool to reach a detailed understanding of the structure of hadrons, and has received a lot of attention in last decades.
In particular, the polarization phenomena of hadrons (proton or lambda hyperon) produced in high energy can provide further information about the spin structure of the hadron and the spin-dependent hadronization mechanism in fragmentation region~\cite{Burkardt:1993zh,Jaffe:1996wp,Ma:2000uv,Anselmino:2001ps,Zhou:2009mx}.
The production of the large transverse polarization of the lambda hyperon in unpolarized $pp$ scattering has been observed~\cite{Bellwied:2002rg,Adler:2002pb,Abelev:2006cs,ATLAS:2014ona} and brought theoretical challenges to the understanding~\cite{Qiu:1991wg,Qiu:1991pp,Liang:2000gz,Anselmino:1994gn,Heller:1978ty,Bunce:1976yb} of those phenomena within Quantum Chromodynamics (QCD).
Recently, the transverse polarization of $\Lambda$ hyperon in $e^+e^-\rightarrow \Lambda^\uparrow+X$ and $e^+e^-\rightarrow \Lambda^\uparrow+h^\pm+X$ was measured by the Belle Collaboration~\cite{Abdesselam:2016nym}, in which the polarized fragmentation function $D^\perp_{1T}$ plays a major part in these effects.
The function $D^{\perp\ h/q}_{1T}$ represents the transverse momentum dependence of a transversely polarized hadron $h$ fragmented from an unpolarized quark $q$.
Besides model calculations~\cite{Metz:2002iz,Yang:2017cwi},
these measurements provides an important approach for future extraction of $D^{\perp\Lambda/q}_{1T}$.
On the other hand, The polarized fragmentation functions may also be used as a probe~\cite{Baldracchini:1980uq} to explore the partonic structure of the nucleon through the semi-inclusive deep inelastic scattering (SIDIS).
The renowned example is the Collins fragmentation function~\cite{Collins:1992kk}, which plays an important role in accessing the transverse spin structure of the nucleon.
Furthermore, in recent years, the production of transversely polarized hadron in SIDIS has been studied both by experiments~~\cite{Rith:2007zz,Ferrero:2007zz,Negrini:2009oia,Airapetian:2014tyc,Karyan:2016rls} and by theory ~\cite{Boer:1997nt,Zhou:2008fb,Yang:2016qsf,Anselmino:2001js,Kanazawa:2015jxa}.

In this work, we extend the phenomenological study of the polarized hadron production in SIDIS at the twist-3 level within the transverse momentum dependent (TMD) framework.
To be clear, the approach is different from the collinear twist-3 formalism which is also applied to study the same process~\cite{Zhou:2008fb,Kanazawa:2015jxa}.
We focus on the $\sin\phi_{S_h}$ azimuthal asymmetry in the transversely polarized hadron production in SIDIS off an unpolarized nucleon: $\ell+N\longrightarrow \ell'+\,h^\uparrow +\,X$, where $\phi_{S_h}$ is the azimuthal angle of the transverse spin $S_{hT}$ with respect to the lepton plane.
In this process, only $\phi_{S_h}$ is need to be measured, and the azimuthal angle of the final-state hadron is integrated out.
We not only consider the case the final hadron is a lambda hyperon, but also include the case the proton is the fragmenting hadron.
Following Ref.~\cite{Mulders:1995dh,Yang:2016qsf}, for the asymmetry in the transversely polarized production of hadron, denoted by $A_{UUT}^{\sin\phi_{S_h}}$, there are several contributions coming from the convolutions of the twist-2 TMD fragmentation functions with twist-3 TMD distribution functions.
Particularly, we investigate the contributions of the $h H_1$, $g^\perp G_{1T}$ and $f^\perp D_{1T}^\perp$ coupling to the SSA $A_{UUT}^{\sin\phi_{S_h}}$ for the proton production as well as the lambda production in SIDIS.
Here $h$, $g^\perp$ and $f^\perp$ are the twist-3 TMD distributions, and $H_1$, $G_{1T}$ and $D_{1T}^\perp$
are the twist-2 polarized TMD fragmentation functions coupled to distributions.
Particularly, $H_1$ is similar to the transversity distribution $h_1$, and $G_{1T}$ describes the transverse momentum dependence of the transversely polarized fragmenting hadron from a longitudinally polarized quark.
To this aim, we calculate the fragmentation functions $H_1$, $G_{1T}$, $D^\perp_{1T}$ of the proton, and  $H_1$, $G_{1T}$ of the lambda hyperon for light flavors, using a spectator model first introduced in Ref.~\cite{Jakob:1997wg}.
The $D^\perp_{1T}$ of the lambda hyperon has already been calculated in Ref.~\cite{Yang:2017cwi} with the same model.
For the twist-3 TMD distributions, the T-odd distribution $g^\perp$ for the $u$ and $d$ quarks was calculated~\cite{Mao:2013waa} in a spectator diquark model by including both the scalar and axial-vector diquarks~\cite{Bacchetta:2008af}, while another T-odd distribution $h$ has only been calculated in a scalar-diquark model~\cite{Lu:2012gu}.
In this work we will calculate $h$ and $f^\perp$ for the $u$ and $d$ quarks using the same spectator diquark model from Ref.~\ref{Bacchetta:2008af}.
Using the model results of those distribution and fragmentation functions, we predict the SSA $A_{UUT}^{\sin\phi_{S_h}}$ in the transversely polarized hadron production in SIDIS at the JLab 12 GeV, with the hadron being a proton or a lambda hyperon. Particularly, We investigate the contributions of the $h H_1$, $g^\perp G_{1T}$ and $f^\perp D_{1T}^\perp$ terms for comparison.
The same asymmetry at the kinematics of COMPASS is also presented.

The remained content of the paper is organized as follows.
In Section II, we calculate the twist-3 TMD distributions $h$, and $f^\perp$ for the $u$ and $d$ valence quarks in a proton using the spectator-diquark model from Ref.~\cite{Bacchetta:2008af}.
In Section III, the twist-2 TMD fragmentation functions $H_1$, $G_{1T}$, $D^\perp_{1T}$ are calculated.
In the Section IV, we present the prediction on the $\sin\phi_{S_h}$ asymmetry at the kinematics of JLab 12 GeV and COMPASS.
Finally, we give our conclusion in Section V.

\section{The calculation of twist-3 TMD distribution functions in a spectator model}

In this section, we perform the calculation of the twist-3 TMD distribution functions $h$, $f^\perp$ in a spectator diquark model following the approach in Ref.~\cite{Bacchetta:2008af}, in which the isospins of the vector diquarks were used to distinguished the
isoscalar (ud-like) spectators and the isovector (uu-like) spectators.
The same model was previously applied to calculate the twist-3 distribution functions and the corresponding azimuthal asymmetries in Refs.~\cite{Mao:2016hdi,Mao:2014fma,Mao:2014dva,Mao:2014aoa,Mao:2013waa,Mao:2012dk}.

The twist-3 TMD distribution functions $h$, $f^\perp$ can be obtained from the quark-quark correlator $\Phi(x,\bm{p_T};S)$ via the following traces,
\begin{align}
\frac{1}{4}\Tr[(\Phi(x,\bm p_T;S)+\Phi(x,\bm {p}_T;-S)) \,i\sigma^{\alpha\beta}\gamma_5]&=-\frac{M}{P^+}\epsilon_T^{\alpha\beta}h,
\label{h}\\
\frac{1}{4}\Tr[(\Phi(x,\bm {p}_T;S)+\Phi(x,\bm {p}_T;-S))\,\gamma^{\alpha}]&=\frac{p_T^{\alpha}}{P^+}f^\perp\,,
\label{fperp}\\
 \frac{1}{4}\Tr[(\Phi(x,\bm {p}_T;S)+\Phi(x,\bm {p}_T;-S))\, \gamma^{\alpha}\gamma_5]&=-\frac{\epsilon_T^{\alpha\rho} p_{T\rho}}{P^+}g^\perp\,.
 \label{gperp}
\end{align}
The TMD quark-quark correlator $\Phi(x,\bm{p_T};S)$ is defined as~\cite{Bacchetta:2008af,Bacchetta:2006tn}
\begin{align}
\Phi(x,\bm {p}_T;S)&=\int {d\xi^- d^2\bm{\xi}_T\over (2\pi)^3}e^{ip\cdot\xi}
\langle P,S|\bar{\psi}(0)\mathcal{L}[0,\xi]\psi(\xi)|P,S\rangle\bigg|_{\xi^+=0}\,,
\label{Phi}
\end{align}
where $p$ and $P$ are the momenta of the quark and the target hadron.
The Wilson line ${\cal U}$ is included to ensure the gauge invariance of the operator, it arises from the gluon exchanges between the active quark and the spectator in hadron~\cite{Bacchetta:2006tn,Bacchetta:2007wc}.

In the spectator diquark model, we can insert a complete set of the intermediate states $|P-p\rangle$ into the correlator in Eq.~(\ref{Phi}) and get the following form
\begin{align}
\Phi^{(0)}(x,\bm {p}_T)=\frac{1}{(2\pi)^3}\,\frac{1}{2(1-x)P^+}\,
\overline{\mathcal{M}}^{(0)}\, \mathcal{M}^{(0)} ,
\label{eq:Phi-tree-spect}
\end{align}
where $\mathcal{M}^{(0)}$ is the nucleon-quark-spectator scattering amplitude at the lowest order
\begin{align}
\mathcal{M}^{(0)}
&=\langle P-p |\psi(0) |P\rangle \nonumber\\
&=\begin{cases}
  \displaystyle{\frac{i}{p\ssh-m}}\, \Upsilon_s \, U(P)\\
  \displaystyle{\frac{i}{p\ssh-m}}\, \varepsilon^*_{\mu}(P-p,\lambda)\,
        \Upsilon_a^{\mu}\, U(P).
  \end{cases}
\label{eq:m-tree}
\end{align}
Here $\varepsilon_{\mu}(P-p,\lambda)$ is the polarization vector of the axial-vector diquark, and the nucleon-quark-diquark vertices $\Upsilon_{s}$ and $\Upsilon_{a}^\mu$ ($s$ for the scalar diquark and $a$ for the axial-vector diquark) can be chosen as the following form~\cite{Jakob:1997wg}
\begin{align}
\Upsilon_s(p^2) = g_s(p^2),~~~
\Upsilon_a^\mu(p^2)={g_a(p^2)\over \sqrt{2}}\gamma^\mu\gamma^5, \label{eq:vertex}
\end{align}
where $g_X(p^2)$ is the dipolar form factor to regularize light-cone divergences in the calculation of T-odd DFs when using a point-like coupling~\cite{Gamberg:2006ru}.
The corresponding expression can be written as
\begin{align}
g_X(p^2)&= N_X {p^2-m^2\over |p^2-\Lambda_X^2|^2}= N_X{(p^2-m^2)(1-x)^2\over
(\bm {p}_T^2+L_X^2)^2},~~ X=s,a, \label{eq:gx}
 \end{align}
with $N_X$ and $\Lambda_X$ the normalization constant and the cut-off parameter of model, respectively, and
$L_X^2$ has the form
\begin{align}
L_X^2=(1-x)\Lambda_{X}^2 +x M_{X}^2-x(1-x)M^2.
\end{align}

Thus, the expression of the quark-quark correlator at tree level contributed by the scalar diquark component is
\begin{align}
\Phi^{(0)}_s(x,\bp)&\equiv \frac{N_s^2(1-x)^3}{32 \pi^3 P^+}\frac{\left[ (p\ssh +m)\gamma_5 S\ssh (P\ssh +M) (p\ssh +m)\right]}{(\bp^2+L_s^2)^4}, \label{eq:lophis}
\end{align}
and by the axial-vector diquark component is
\begin{align}
 \Phi^{(0)}_{a}(x,\bp)&\equiv \frac{N_a^2(1-x)^3}{64 \pi^3 P^+}d_{\mu\nu}(P-p)\frac{\left[(p\ssh +m)\gamma^{\mu}\gamma_5 S\ssh(M-P\ssh )\gamma^{\nu}(p\ssh+m)\right]}{(\bp^2+L_a^2)^4}.
\label{eq:lophiv}
\end{align}
In Eqs.~(\ref{eq:lophis}) and (\ref{eq:lophiv}), $p^+ = xP^+$ and the summation is over the polarizations of the axial-vector diquark
$d_{\mu\nu}(P-p)=\Sigma_{\lambda} \varepsilon^{*}_{\mu}(\lambda)\varepsilon_{\nu}(\lambda)$.
In this work, we adopt the following form for the propagator $d_{\mu\nu}$
\begin{align}
d_{\mu\nu}(P-p)  =\,-g_{\mu\nu}\,+\, {(P-p)_\mu n_{-\nu}
 \,+ \,(P-p)_\nu n_{-\mu}\over(P-p)\cdot n_-} - \,{M_a^2 \over\left[(P-p)\cdot n_-\right]^2 }\,n_{-\mu} n_{-\nu}.\label{eq:polarizations}
\end{align}

Following Ref.~\cite{Mao:2014dva}, we can calculate the T-even distribution $f^\perp$ in the spectator model, by considering both the scalar and the axial-vector diquark.
Since $f^\perp$ is a T-even distribution, one can obtain its expression from the lowest-order correlator.
However, $g^\perp$ and $h$ are T-odd distributions and vanish in the lowest order.
The nonzero results for T-odd distributions arise from the effect of the gauge link~\cite{Brodsky:2002cx,jy02,Collins:2002plb}.
Thus, we have to compute the one-loop amplitude to obtain the imaginary part.
In the spectator model, the interference of the lowest-order amplitude $\mathcal M^{(0)}$ and the one-loop-order amplitude $\mathcal M^{(1)}$ gives rise to following contribution to the correlator
\begin{align}
  \Phi_s^{(1)}
(x,\bm {p}_T)
&\equiv
-i e_q N_{s}^2 {(1-x)^2\over 64\pi^3 (P^+)^2}\frac{-i\Gamma^{+}_s}{(\bm{p}_T^2+L_s^2)^2}\nonumber \\
\hspace{-1cm}&\times \int {d^2 \bm q_T\over (2\pi)^2}
{ \left[(\pslash -q\ssh+m)(\Pslash+M)(\pslash +m)\right]
\over \bm q_{T}^2  \left[(\bm{p}_T-\bm{q}_T)^2+L_s^2\right]^2}
 , \label{phis1}\\
 \Phi^{(1)}_{a}
(x,\bm {p}_T)
&\equiv
-i e_q N_a^2 {(1-x)^2\over 128\pi^3 (P^+)^2}{1\over (\bm{p}_T^2+L_a^2)^2}\nonumber\\
&\times\int {d^2 \bm q_T\over (2\pi)^2} \,
 d_{\rho\alpha}(P-p)\, (-i\Gamma^{+,\alpha\beta}) \nonumber\\
 &\times d_{\sigma\beta}(P-p+q) \nonumber\\
&\times{ \left[(\pslash -q\ssh+m) \gamma^\sigma(\Pslash-M)\gamma^\rho (\pslash +m)\right]
\over \bm q_T^2  \left[(\bm{p}_T-\bm{q}_T)^2+L_a^2\right]^2},
\label{phia1}
\end{align}
with $q^+=0$.

The vertex between the gluon and the scalar ($\Gamma_s$) or axial-vector diquark ($\Gamma_a$) has the following form
 \begin{align}
 \Gamma_s^\mu &= ie_s (2P-2p+q)^\mu, \\
 \Gamma_a^{\mu,\alpha\beta} &=  -i e_a [(2P-2p+q)^\mu g^{\alpha\beta}-(P-p+q)^{\alpha}g^{\mu\beta}-(P-p)^\beta g^{\mu\alpha}]\label{Gamma}.
\end{align}
After Performing the trace, we can get the corresponding expressions from the scalar component to $h$, $f^\perp$ and $g^\perp$:
\begin{align}
h(x,\bm{p}_T^2)_s=&-{ N_s^2(1-x)^3\over 16\pi^3 M}{e_se_q\over 4\pi}{(m+xM)(L_s^2-\bm{p}_T^2)\over  L_s^2(\bm{p}_T^2+L_s^2)^3}\,,\\
f^\perp(x,\bm{p}_T^2)_s=&-{N_s^2(1-x)^2\over 16\pi^3 }{(\bm {p}_T^2-2mM(1-x)-(1-x^2)M^2+m_s^2)\over (\bm {p}_T^2+L_s^2)^4}\,.
\label{scalar}
\end{align}
Similar to scalar diquark case, we obtain the final results from the axial-vector diquark component:
\begin{align}
h(x,\bm {p}_T^2)_a=&0\,,\\
f^\perp(x,\bm {p}_T^2)_a=&{N_a^2(1-x)\over 16\pi^3 }{(x\bm {p}_T^2+2mM(1-x)^2+(x-1)m^2+(x^3-2x^2+1)M^2-m_a^2)\over  L_a^2(\bm {p}_T^2+L_a^2)^3}\,.
\end{align}

In order to obtain the distribution functions of the u and d quarks with $f^{s}$ and $f^{a}$, we use the following relation between the flavors and isospins of the diquark~\cite{Bacchetta:2008af}
\begin{align}
f^u=c_s^2 f^s + c_a^2 f^a,~~~~f^d=c_{a^\prime}^2 f^{a^\prime}\label{ud},
\end{align}
where $c_s$, $c_a$ and $c_{a^\prime}$ are the free parameters of the model, $a$ and $a^\prime$ denote the isoscalar and isovector states of the axial-diquark, respectively.
These parameters, together with the mass parameters (such as the diquark masses $M_X$, the cut-off parameters $\Lambda_X$), are taken from Ref.~\cite{Bacchetta:2008af}.
In this work, we use the following replacement for the combination of the charges of the quark and the spectator diquark
\begin{align}
{e_qe_X\over 4\pi} \rightarrow -C_F \alpha_s,
\end{align}
and we choose $\alpha_s~\approx 0.3$ in our calculation.

\begin{figure}
  \includegraphics[width=0.49\columnwidth]{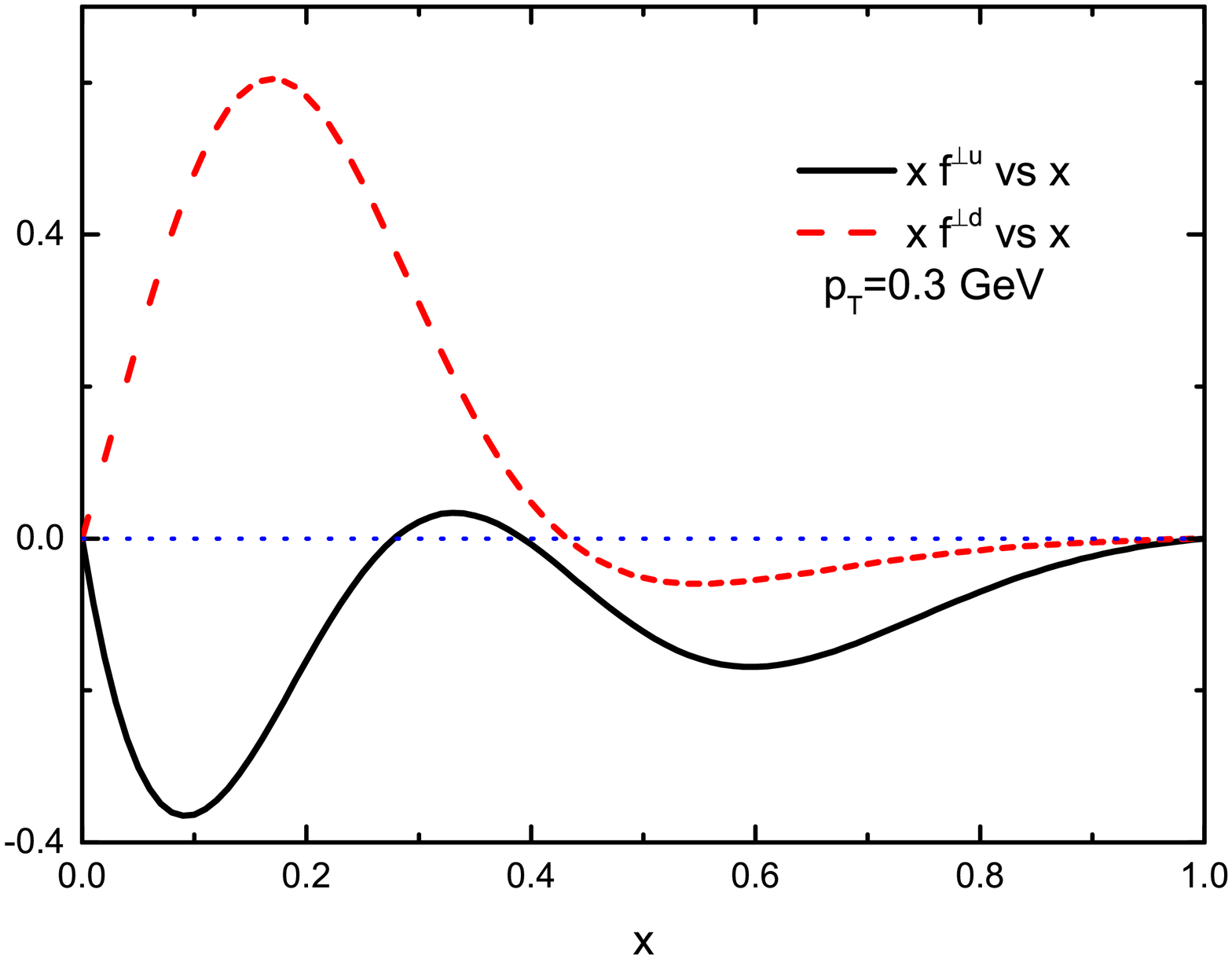}
  \includegraphics[width=0.49\columnwidth]{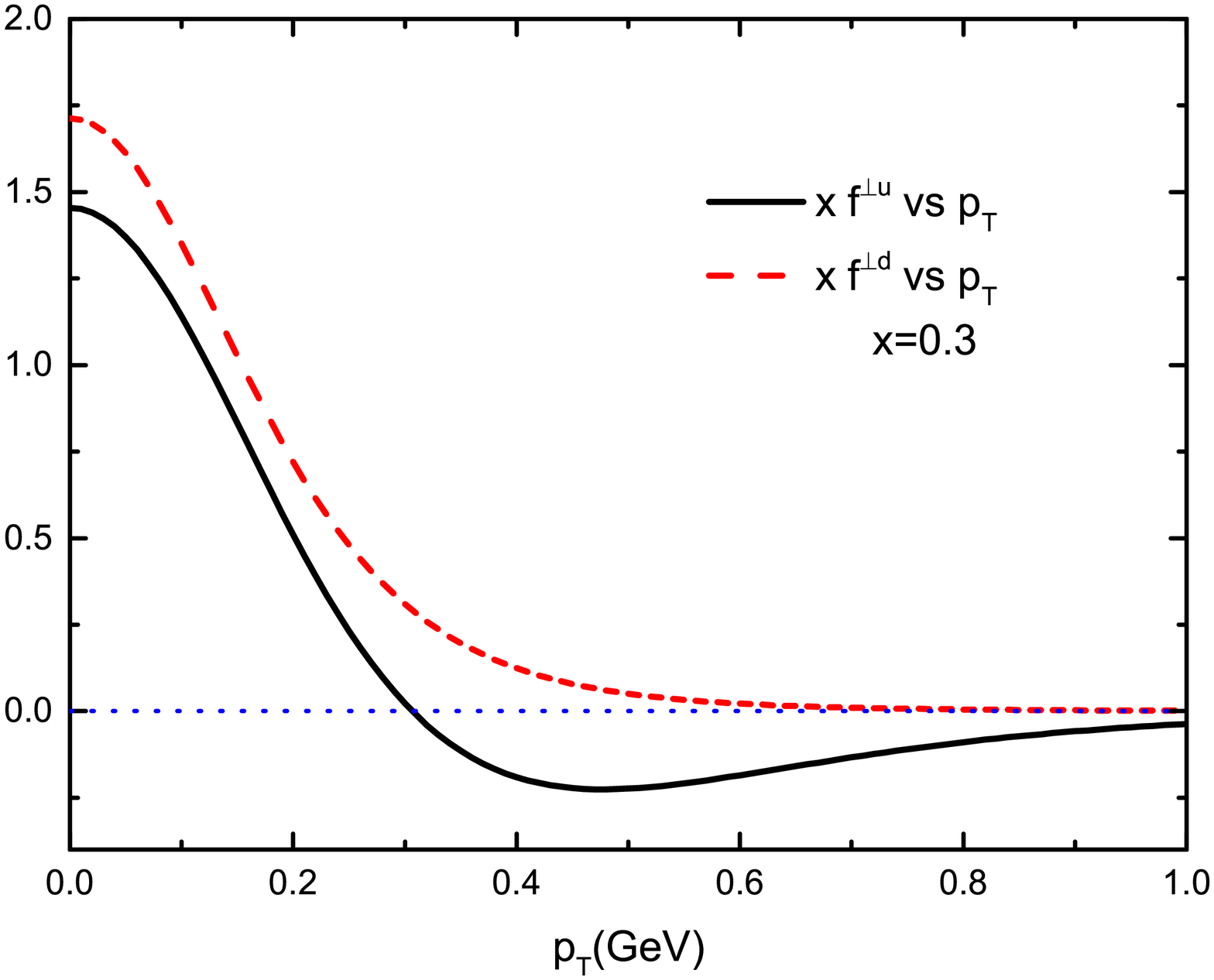}
  \caption{Left panel: model results for $x f^{\perp u}(x,\bp^2)$ (solid line) and $x f^{\perp d}(x,\bp^2)$ (dashed line) as functions of $x$ at $p_T=0.3\,\text{GeV}$. Right panel: model results for $x f^{\perp u}$ (solid line) and $x f^{\perp d}$ (dashed line) as functions of $p_T$ at $x=0.3$.}\label{FIG:fperp}
\end{figure}
\begin{figure}
  \includegraphics[width=0.49\columnwidth]{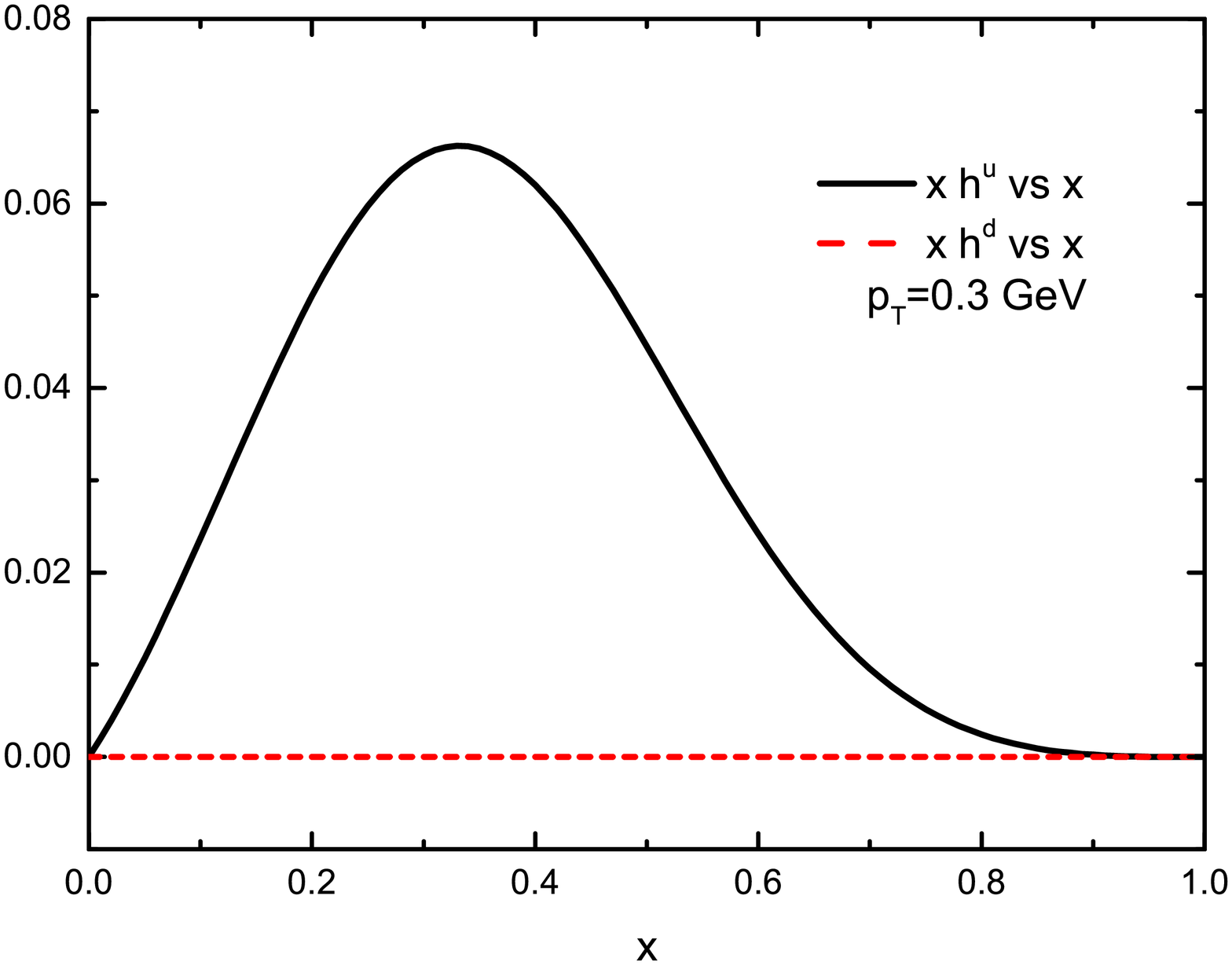}
  \includegraphics[width=0.49\columnwidth]{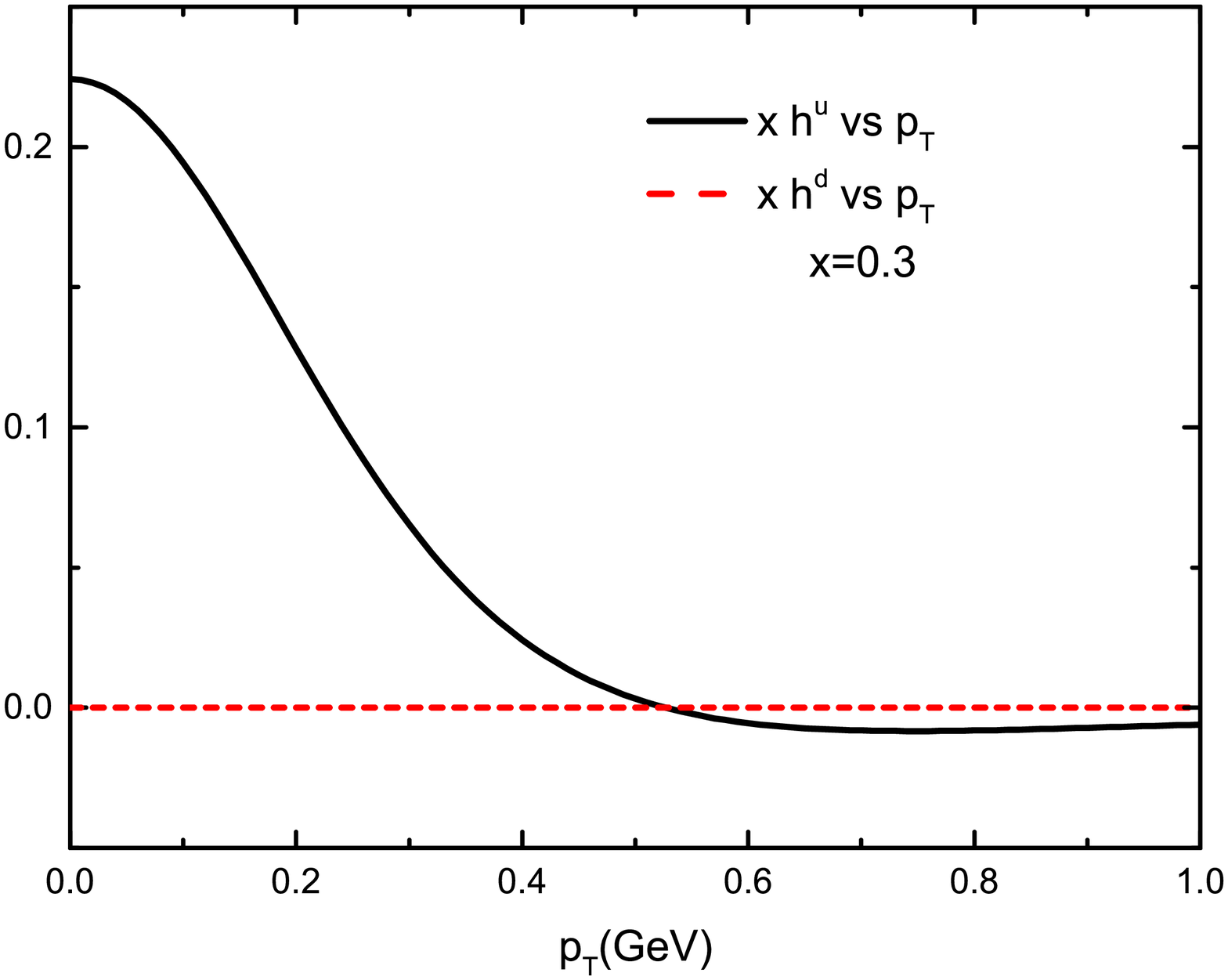}
  \caption{Similar to Fig.~\ref{FIG:fperp}, but for the model results of $x h^{u}(x,\bp^2)$ (solid line) and $x h^{d}(x,\bp^2)$ (dashed line).}\label{FIG:h}
\end{figure}

In the left panel of Fig.~\ref{FIG:fperp}, we plot the T-even distribution $f^\perp$ as function of $x$ at a fixed $p_T=0.3 GeV$, and in the right panel we plot $f^\perp$ vs $p_T$ at x=0.3.
The solid and lines show the results for the u and d valence quark, respectively.
As we can see, in the specified kinematic region ($x=0.3$ or $p_T=0.25$ GeV), the distributions $f^{\perp\,u}$ and $f^{\perp\,d}$ are in similar sizes.

In Fig.~\ref{FIG:h}, we show the curves of the T-odd distribution $h$.
Since the axial-vector diquark contribution $h^a$ vanishes in the model, $h^d$ is zero.
We find that at low $p_T$, $h^u$ is positive, while it is negative in the intermediate range of $p_T$, and eventually falls to zero at large $p_T$. That is, there is a node of the distribution $h^u$ in $p_T$.
The size of $h$ is smaller compared to those of the T-even distributions $f^\perp$.
In particular, with the $p_T$-dependence of $h$ given in Eq.~(\ref{scalar}), we can verify that $h^u$ vanishes when one integrates out the transverse momentum $\bp$ ~\cite{Goeke:2005hb}
\begin{align}
\int d^2\bp h^u(x,\bm {p}_T^2)=0.
\end{align}
This is an expected result from the time-reversal invariance for integrated distributions, and it also indicates that the distribution $h$ will not give any contribution to the transverse SSA in inclusive DIS process~\cite{Metz:2012ui,Airapetian:2009ab}.

\section{The calculation of TMD fragmentation function in spectator model}

In this section, we calculate the three twist-2 TMD fragmentation functions $H_1$, $G_{1T}$ and $D^\perp_{1T}$ that corresponding to the transverse polarization hadron production in a spectator model~\cite{Jakob:1997wg}.
We note that, besides the model calculation, independent information of $H_1$, $G_{1T}$ and $D^\perp_{1T}$ may be also accessible in the electron-positron annihilation process $e^+ e^- \to h_1^\uparrow h_2^\uparrow X$~\cite{Abdesselam:2016nym,Boer:1997mf}.

Similar to the calculation on the distribution functions, the TMD polarized fragmentation functions $H_1$, $G_{1T}$ and $D^\perp_{1T}$ may be obtained from the fragmentation correlation function $\Delta(z,\bm{k}_T;\bm{S}_{h\,T})$ by the following trace
\begin{align}
{\epsilon_T^{\alpha\beta}k_{T\alpha}S_{hT\beta}\over M} D^\perp_{1T}(z,\bm {k}_T^2) = &{1\over 2}\textrm{Tr}[(\Delta(z,\bm{k}_T;\bm{S}_{h\,T})-\Delta(z,\bm{k}_T;-\bm{S}_{h\,T}))\gamma^-]\,,\\
S_{h{L}}\,G_{1L}(z,\bm {k}_T^2)+{\bm{k}_T\cdot\,\bm{S}_{hT}\over M_h }G_{1T}(z,\bm {k}_T^2)=&{1\over 4}\textrm{Tr}[(\Delta(z,\bm{k}_T;\bm{S}_{h\,T})-\Delta(z,\bm{k}_T;-\bm{S}_{h\,T}))  \gamma^-\gamma_5]\,,\\
S^\alpha_T\,H_1(z,\bm k_T^2) =& {1\over 4}\textrm{Tr}[(\Delta(z,\bm{k}_T;\bm{S}_{h\,T})-\Delta(z,\bm{k}_T;-\bm{S}_{h\,T})) i\sigma^{\alpha\,-}\gamma_5]\,,
\end{align}
with
\begin{align}
\Delta(z,\bm{k}_T;\bm{S}_{h\,T})=\frac{1}{2z}\sum_X \, \int{d\xi^+d^2\bm{\xi_T}\over(2\pi)^3} e^{ik\cdot\,\xi}\langle 0|\, {\cal U}^{n^+}_{(+{\infty},\xi)}
\,\psi(\xi)|P_h,S_h; X\rangle\langle P_h,S_h; X|\bar{\psi}(0)\,
{\cal U}^{n^+}_{(0,+{\infty})}
|0\rangle \bigg|_{\xi^-=0}\,,
\label{eq:delta}
\end{align}
where where the final state $|P_h,S_h; X\rangle$ describes the final state hadron and the intermediate unobserved states.
The spin vector $S_h$ of the outgoing hadron is decomposed as
\begin{align}\label{spin}
S_{h}^\mu = S_{hL}{(P_h\cdot n_+)n_-^\mu\,- (P_h\cdot n_-)n_+^\mu\over M_h}+ S_{hT}^\mu\,.
\end{align}

In this section, we chose the forms of the vertex and propagator $d_{\mu\nu}$ for fragmentation functions from Ref.~\cite{Jakob:1997wg}.
The choice has also been applied to calculate the fragmentation functions $D^\perp_{1T}$ and $H^\perp_1$ of the $\Lambda$ hyperon in Refs.~\cite{Yang:2017cwi,Wang:2018wqo}.
We also give the expression of the matrix element
\begin{align}\label{eq:diquark}
  \langle\,P_h,S_h; X|\,\bar\psi(0)|0\rangle=
  \begin{cases}
  \bar{U}(P_h,S_h)\, {\mathcal{Y}}_s\,\displaystyle{\frac{i}{\kslash-m_q}}&     \textrm{scalar diquark,} \\
 \bar{U}(P_h,S_h\,){\mathcal{Y}}^{\mu}_v \,\displaystyle{\frac{i}{\kslash-m_q}}\,\varepsilon_{\mu}\, & \textrm{axial-vector diquark.}
  \end{cases}
\end{align}
Here $\mathcal{Y}_{D}$ ($D=s$ or $v$) is the hyperon-quark-diquark vertex, and $\varepsilon_{\mu}$ is the polarization vector of the spin-1 vector diquark.
The summation for all polarizations states of the vector diquark: $\sum_\lambda\varepsilon^{*(\lambda)}_\mu\varepsilon^{(\lambda)}_\nu=-g_{\mu\nu}
+{P_{h\mu}\,P_{h\nu}\over M_h^2}$.
Besides, in this work, the vertex structure in fragmentation is chosen as follows~\cite{Jakob:1997wg}
\begin{align}
  \mathcal{Y}_s&=\bm{1}g_s\,,\notag\\
  \mathcal{Y}^\mu_v &={g_v\over \sqrt{3}}\gamma_5(\gamma^\mu+{P_h^\mu\over M_h})\,,
\end{align}
where $g_D$ ($D=s$ or $v$) is the suitable coupling for the baryon-quark-diquark vertex.
In this paper, the coupling vertex $g_D$ is chosen as a $k^2$ dependent Gaussian form factor to cut the divergence from the large $k_T$ region:
\begin{align}\label{factor}
  g_D~\mapsto {g_D\over z}\,e^{-{k^2\over \Lambda^2}}\,,
\end{align}
where $\Lambda^2$ has the general form $\Lambda^2= \lambda^2z^\alpha(1-z)^\beta$.
The parameters of the model are $\lambda, \alpha, \beta$, together with the masses of the spectator diquark $m_D$ and the parent quark $m_q$.
Using the above settings, we can gain the theoretical expressions of $D_1$ as follows,
\begin{align}\label{D1}
 D^{(s)}_1(z,\bm k_T^2) = D^{v}_1(z,\bm k_T^2) = {g_D^2\over 2(2\pi)^3}{1 \over z^2}e^{-2k^2\over\Lambda^2}{(1-z)[z^2k^2_T+(M_\Lambda+zm)^2]\over \,z^4(k_T^2+L^2)^2}\,,
\end{align}
 which has already been given in Ref.~\cite{Yang:2017cwi}.

We note that our choice for polarization sum and the vertex structure of the vector diquark in fragmentation process is different from the one used in the calculation the TMD distributions shown in the previous Section.
The difference comes from two folds. Firstly, as the fragmentation mechanism is different from the distribution of quark in a hadron, the spectator states in fragmentation are not necessarily the same as those in the initial hadron.
Secondly, we find that with the current choice, our model result can reasonably reproduce the unpolarized proton fragmentation function agreeing with the available parametrization~\cite{Hirai:2007cx}.

\begin{figure}
  \includegraphics[width=0.49\columnwidth]{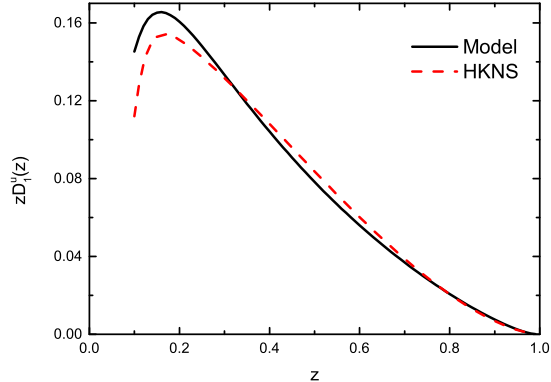}
  \caption{Unpolarized fragmentation function $zD_1^u(z)$ of the proton (solid line) compared with the HKNS parametrization (dashed line) is shown for comparison.}
  \label{FIG:Ff}
\end{figure}

Assuming the SU(6) spin-flavor symmetry for the final state hadron~\cite{Hwang:2016ikf,VanRoyen:1967nq,Jakob:1993th},
We can write the relation between quark flavors and diquark types for the proton and the lambda hyperon as \begin{align}
D^{\text{u}\rightarrow p}&={3\over 2}D^{(s)}+{1\over 2} D^{(v)},~~~ D^{\text{d}\rightarrow p}=D^{(v)}\,, D^{\text{s}\rightarrow p}=0\label{eq:Dproton} \\
D^{\textrm{u}\rightarrow \Lambda} &=\,D^{\textrm{d}\rightarrow \Lambda} ={1\over 4}D^{(s)}+{3\over 4}D^{(v)}\,,~~D^{\textrm{s}\rightarrow \Lambda}=D^{(s)}\,,\label{eq:Dlambda}
\end{align}
where $\text{u}$, $\text{d}$, $\text{s}$ denote the up, down and strange quarks, respectively.
According to Eq.~(\ref{eq:Dproton}) and Eq.~(\ref{D1}), one can find that the unpolarized fragmentation function $D_1$ for proton satisfies the relation: $D^{\text{u}\rightarrow p}_{1}=2D^{\text{d}\rightarrow p}_{1}$.
This result is consistent with the HKNS parametrization of $D_1^p$ for $\text{u}$, $\text{d}$ quarks presented in Ref.~\cite{Hirai:2007cx}.

As we have already presented the result for the lambda fragmentation function $D_1^{\Lambda/q}$ in Ref.~\cite{Yang:2017cwi}, in the following we focus on the proton fragmentation function $D_1^{p/q}$.
In order to get the numerical results, we choose the constituent quark mass as $m= 0.3\,\text{GeV}$ for the up and down quarks, and the proton mass as $0.938\,\text{GeV}$.
For the values of the other parameters, we fit our model expression of $D^p_1(z)$ to the leading order (LO) HKNS parametrization~\cite{Hirai:2007cx} at the initial scale $\mu^2_{LO}=1~\textrm{GeV}^2$.
The fitted values of the parameters are
\begin{align}\label{paramters}
      g_D=1.588^{+0.1}_{-0.096},~~~ m_D=0.849^{+0.04}_{-0.0376}\textrm{GeV},~~~ \lambda=10.192^{+1.34}_{-1.11}\textrm{GeV},~~~ \alpha=0.5(\textrm{fixed}),~~~ \beta=0(\textrm{fixed}).
     \end{align}
Here, the parameters $\alpha$, $\beta$ are fixed in our fit, $m_D$ is the mass of the diquark which is the same for the scalar diquark and the vector diquark in our model. The errors of the parameters corresponds to the assumed $30\%$ uncertainties in the HKNS parametrization.
In Fig.~\ref{FIG:Ff}, we plot our model calculation of the unpolarized fragmentation function $zD^{u\rightarrow p}_1(z)$ (solid line), using the parameters in Eq.~(\ref{paramters}).
The parametrization of the HNKS~\cite{Hirai:2007cx} is also shown for comparison (dashed line).

In Ref.~\cite{Yang:2017cwi}, the T-odd fragmentation function $D^\perp_{1T}$ (for the lambda hyperon) have been calculated by the same diquark spectator model. Here we will use the expression directly from Ref.~\cite{Yang:2017cwi}.
For the rest relevant TMD fragmentation functions $H_1$ and $G_{1T}$, we obtain the spectator model results:
\begin{align}
G^R_{1T}(z,\bm k_T^2)&=a_R{g_D^2\over (2\pi)^3}{1 \over z^2}e^{-2k^2\over\Lambda^2}{M_h(zm+M_h)(1-z)\over \,z^5(\bm k_T^2+L_f^2)^2}\,,\label{eq:G1T}\\
H^R_{1}(z,\bm k_T^2)&=a_R{g_D^2\over 2(2\pi)^3}{1 \over z^2}e^{-2k^2\over\Lambda^2}{(1-z)[(zm+M_h)^2]\over \,z^4(\bm k_T^2+L_f^2)^2}\,,\label{eq:H1}
\end{align}
where
\begin{align}\label{L2}
L_f^2={1-z\over z^2}M_h^2+m^2+{m_D^2-m^2\over z}\,.
\end{align}
The spin factor $a_R$ takes the values $a_s=1$ and $a_v=-{1\over 3}$.

From Eqs.~(\ref{eq:Dproton},\ref{eq:Dlambda}) and (\ref{eq:G1T},\ref{eq:H1}), we find that for the fragmentation function $G_{1T}$ and $H_1$, both the u and d quark fragmenting to the lambda hyperon vanishes in this model, while the u and d quark fragmenting to the proton is nonzero.
Therefore in the next section, we will only consider the contribution from $D_{1T}^\perp$ for the lambda production.

\begin{figure}
  \includegraphics[width=0.49\columnwidth]{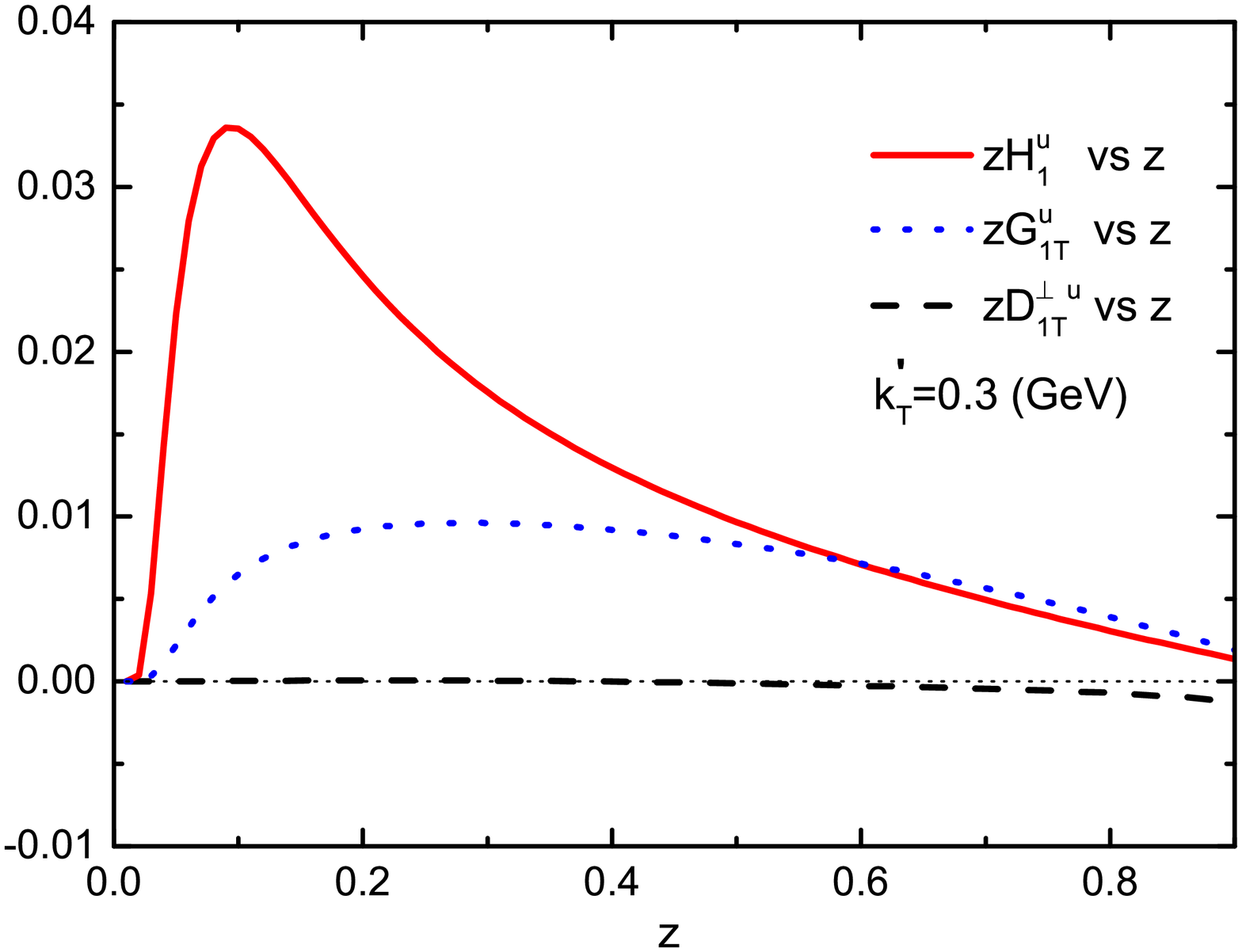}
  \includegraphics[width=0.49\columnwidth]{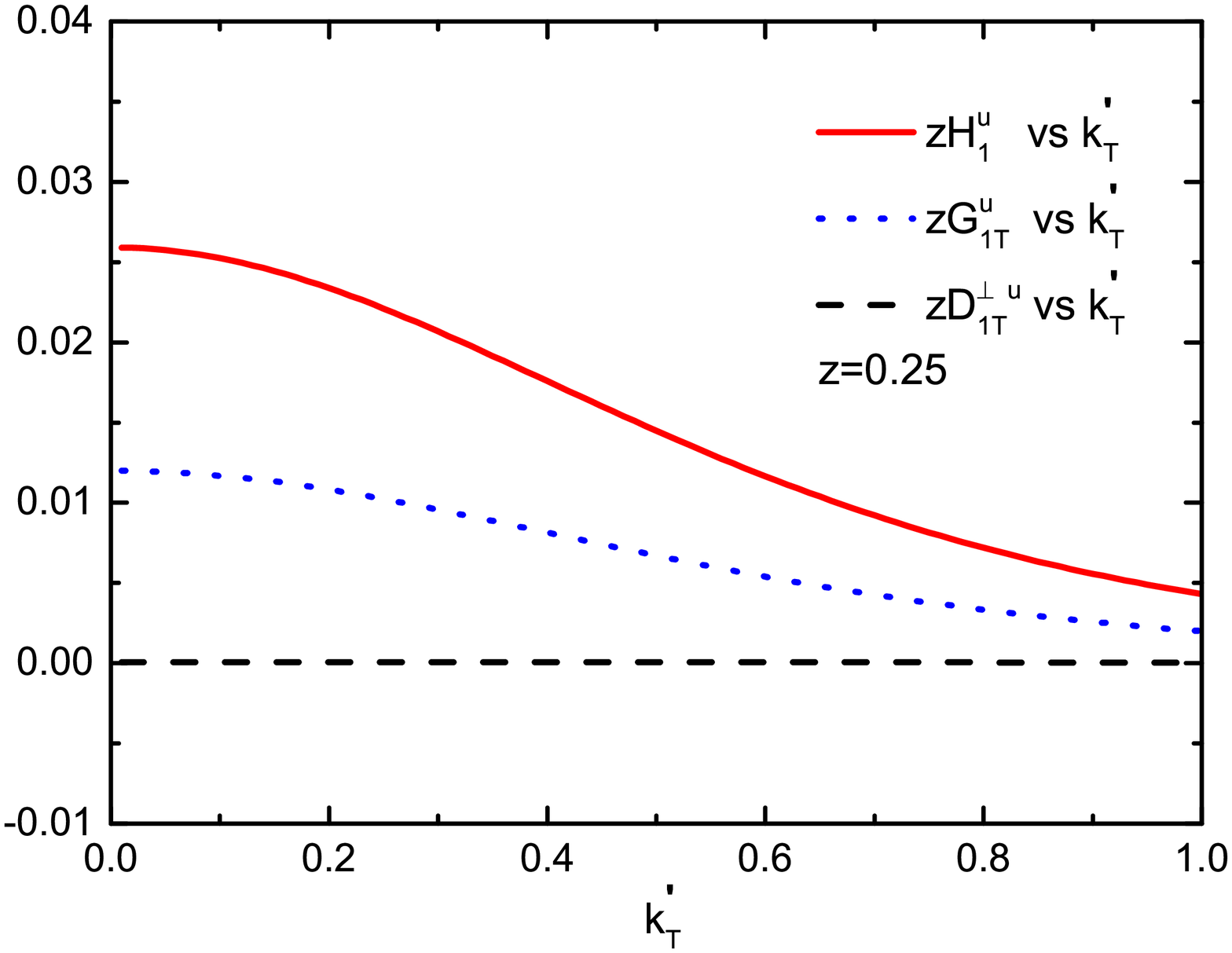}
  \caption{Left panel: the model results of $z H_1^{u}$ (solid line), $zG^u_{1T}$ (dotted line)  and $zD_{1T}^{\perp u}$ (dashed line) for the proton as functions of $z$ at $k_T^\prime=0.3\,\text{GeV}$; Right panel: the model results of $z H_1^{u}$ (solid line), $zG^u_{1T}$ (dotted line) and $zD_{1T}^{\perp u}$ (dashed line) for the proton as functions of $k_T^\prime$ at $z=0.25$.}\label{FIG:uF}
\end{figure}
\begin{figure}
  \includegraphics[width=0.49\columnwidth]{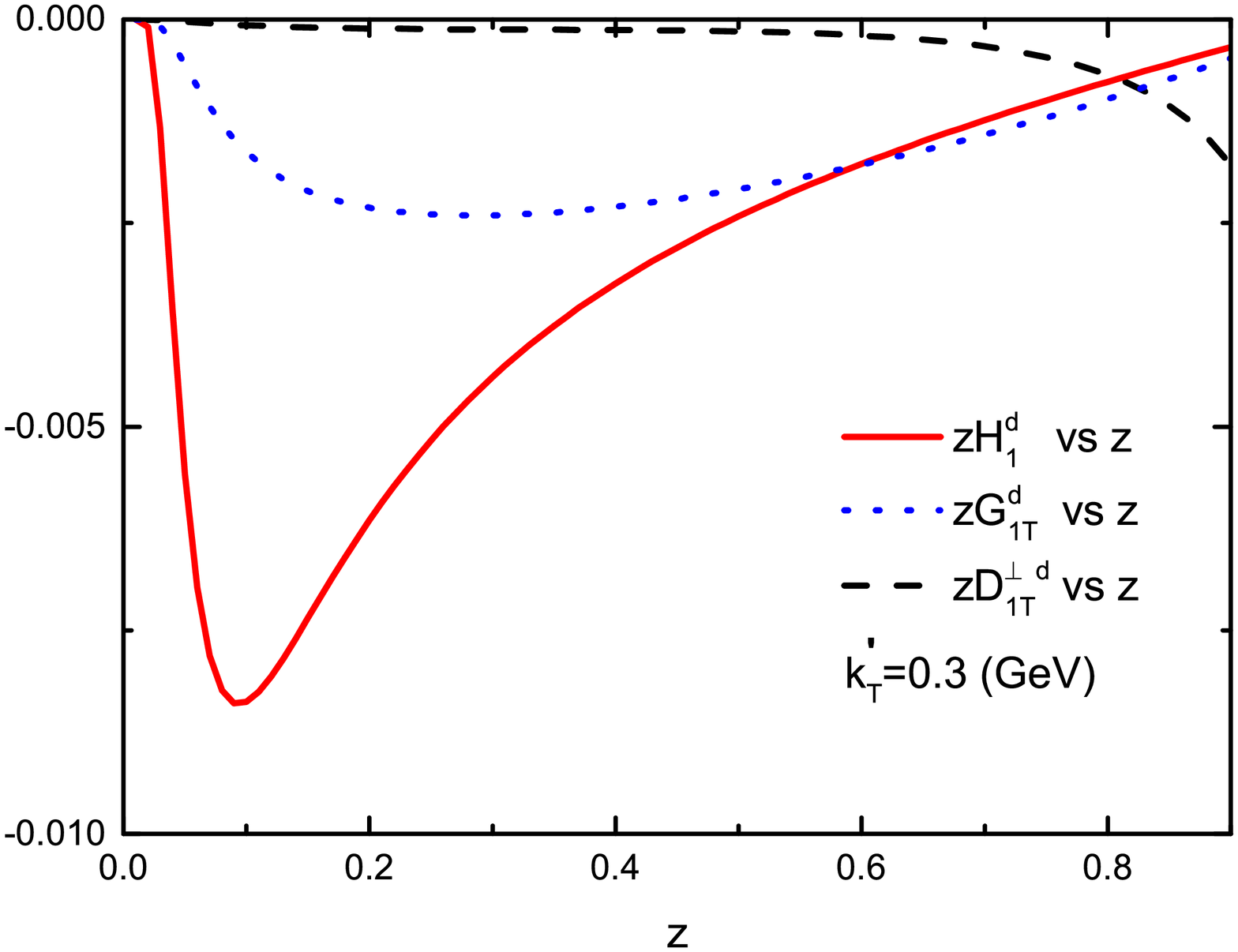}
  \includegraphics[width=0.49\columnwidth]{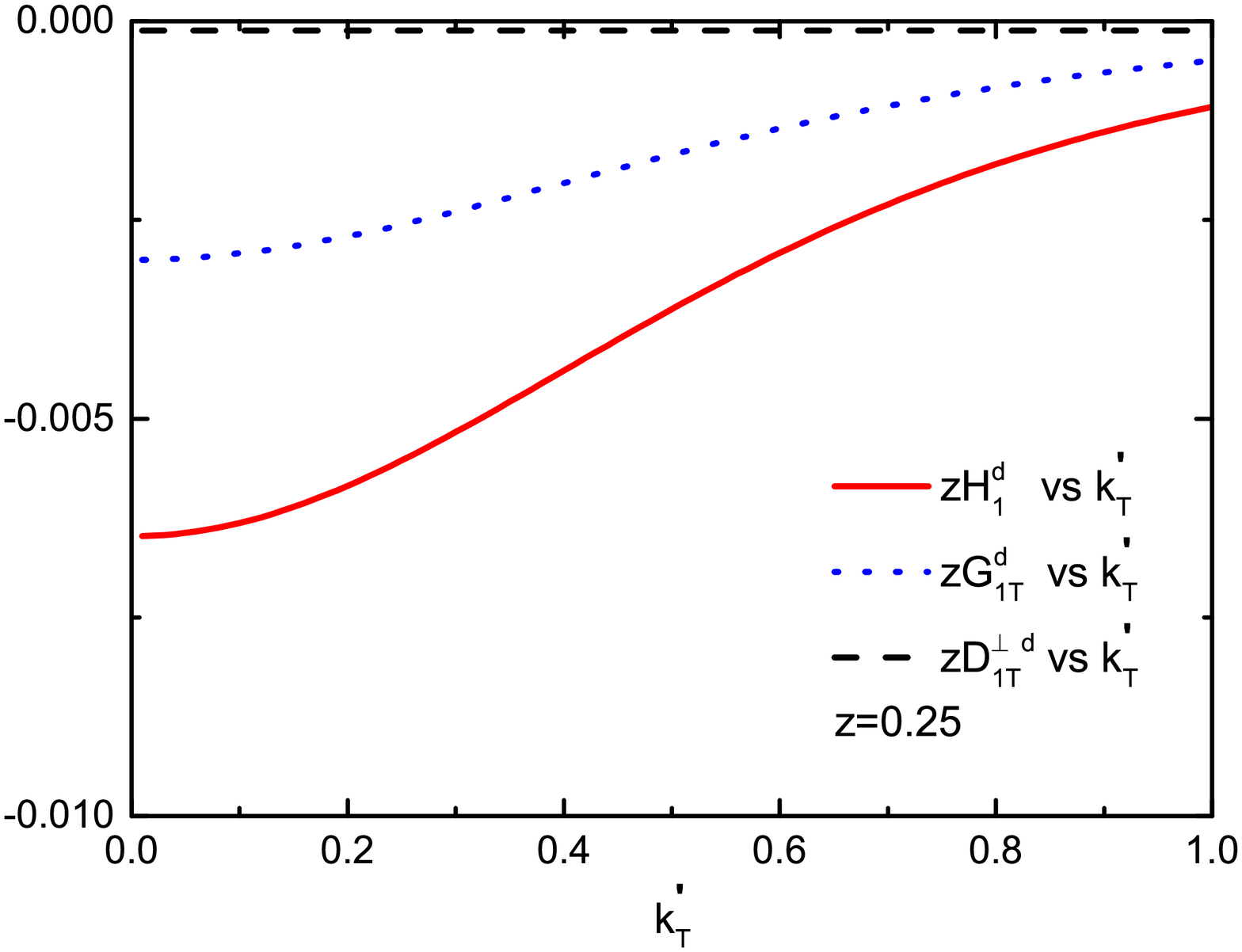}
  \caption{Similar to Fig.~\ref{FIG:uF}, but for the model results of $zH_{1}^{d}$ (solid line), $zG_{1T}^{d}$ (dotted line) and $zD_{1T}^{\perp d}$  (dotted line).}\label{FIG:dF}
\end{figure}

In the left panel of Fig.~\ref{FIG:uF}, we plot the twist-2 TMD fragmentation functions $H_1$, $G_{1T}$ and $D^\perp_{1T}$ for the u quark fragmenting to the proton as functions of $z$ at $k_T^\prime =0.3 GeV$.
Here $k_T^\prime = |\bm k_T^\prime|$, with $\bm k_T^\prime =-z\bm k_T$ representing the transverse momentum of the final hadron with respect to the momentum direction of the parent quark.
In the right panel of Fig.~\ref{FIG:uF}, we plot the same fragmentation functions vs $k_T^\prime$ at $z=0.25$.
We also show the twist-2 TMD fragmentation functions $H_1$, $G_{1T}$ and $D^\perp_{1T}$ of the proton for the d quark as functions of $z$ and $k_T^\prime$ in Fig.~\ref{FIG:dF}.
As we can see, the magnitude of the transversity fragmentation function $H_1$ is lager than that of $G_{1T}$, and $D^\perp_{1T}$ is almost negligible in the small $z$ region.
Further more, $H_1$ and $G_{1T}$ decrease with increasing $z$, while $D^\perp_{1T}$ increases with increasing $z$.
About the sign of the fragmentation functions, $H_1^{u}$ and $G_{1T}^u$ are positive in the model, while $H_1^{d}$ and $G_{1T}^d$ are negative.
Finally, For the T-odd fragmentation function $D^{\perp}_{1T}$ of the proton, only the one for the d quark  becomes sizable at large $z$.

\section{Prediction on the transverse SSA for the productions of proton and lambda hyperon in SIDIS}

The process we are going to study is the production of the transverse polarized hadron in SIDIS off the unpolarized proton:
\begin{align}
l (\ell) \, + \, p (P) \, \rightarrow \, l' (\ell')
\, + \, h^\uparrow(P_h) \, + \, X (P_X)\,,
\label{sidis}
\end{align}
where $\uparrow$ denotes that the final hadron is transverse polarized, $\ell$ and $\ell'$ represent the momenta of the incoming and outgoing leptons, and $P$ and $P_h$ are the momenta of the target nucleon and the final-state hadron, which can be a proton or a lambda hyperon.
\begin{figure}
  \includegraphics[width=0.4\columnwidth]{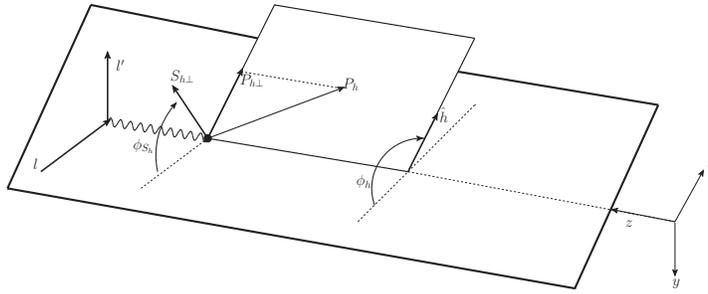}
 \caption {The kinematical configuration for the polarized SIDIS process. The initial and scattered leptonic momenta define the lepton plane ($x-z$ plane), while the detected hadron momentum together with the $z$ axis identify the hadron production plane.}
 \label{SIDISframe}
\end{figure}

Following the convention in Ref.~\cite{Bacchetta:2004jz}, in our calculation we adopt the reference frame shown in Fig.~\ref{SIDISframe}~\cite{Yang:2016qsf}, where $P_{h\perp}$ and ${S}_{h\perp}$ are the transverse momentum and the transverse spin of the detected hadron, respectively. The corresponding azimuthal angles with respect to the lepton scattering plane are denoted by $\phi_h$ and $\phi_{S_h}$.
The invariant variables used to express the differential cross section of SIDIS are defined as
\begin{align}
&x = \frac{Q^2}{2\,P\cdot q},~~~
y = \frac{P \cdot q}{P \cdot l},~~~
z = \frac{P \cdot P_h}{P\cdot q},~~~
\gamma={2M x\over Q},~~~\nonumber\\
&Q^2=-q^2, ~~~
s=(P+\ell)^2,~~~
W^2=(P+q)^2,~~~
\end{align}
where $q=\ell-\ell'$ is the four-momentum of the virtual photon, and $W$ is the invariant mass of the hadronic final state.
With the above variables, the differential cross section of the process (\ref{sidis}) for polarized proton and Lambda hyperon in semi-inclusive deep inelastic scattering off an unpolarized target can be expressed as~\cite{Boer:1997nt,Yang:2016qsf}
\begin{align}
&{{d\sigma}\over{dx dy dz d\phi d\psi dP^2_{h\perp}}}={\alpha^2\over xyQ^2}{y^2\over 2(1-\varepsilon)}(1+{\gamma^2\over 2x})\bigg{\{}F_{UUU}\notag\\
&+|S_{hT}|\sin\phi_{S_{h}}{\sqrt{2\varepsilon(1+\varepsilon)}}F_{UUT}^{\sin\phi_{S_h}}
+\cdots \bigg{\}}
\label{product}
\end{align}
with $F_{ABC}=F_{ABC}(x,z,P^2_{h\perp})$. The subscripts $A, B$ and $C$ indicate the polarizations of the incoming lepton, the target nucleon and the produced final state hadron, respectively.
$F_{\mathrm{UUU}}$ is the spin-averaged structure function, and $F_{\mathrm{UUT}}^{\sin{\phi_S}}$ is the spin-dependent structure function that contributes to the $\sin\phi_{S_h}$ azimuthal asymmetry.

As well known, the structure functions shown in Eq.~(\ref{product}) can be expressed by the convolution of twist-2 and twist-3 TMD distribution and fragmentation.
The two structure functions in Eq.~(\ref{product}) are given by the following expressions ~\cite{Yang:2017cwi}
\begin{align}
F_{UUU}&=\mathcal{I}[f_1D_1]\,,\\
F^{\sin{\phi_{S_h}}}_{UUT}&={2M\over Q}\mathcal{I}\bigg{\{}({M_h\over M}f_1\,{\tilde{D}_T\over z}-x{h\,H_{1}})+{\bm{k}_T\,\cdot\bm{p}_T \over 2MM_h}[({M_h\over M}h_1^\perp\tilde{H}_T^\perp-xf_1\,D^\perp_{1T})-({M_h\over M}h_1^\perp{\tilde{H}_{T}\over z}+xg^\perp\,G_{1T})]\bigg{\}}\,,
\end{align}
where we introduce the convolution integral
\begin{align}
\mathcal{I}\big{[}\omega\,f\,D\big{]}=x\sum_q e_q^2\int d^2 \bm p_T\int d^2\bm k_T\delta^2(\bm p_T-{\bm P_{h\perp}\over z}-\bm k_T) w(\bm p_T, \bm k_T)f^q(x,\bm p_T^2) D^q(z,\bm k_T^2)\,.
\end{align}
Based on the Wandzur-Wilczek approximation~\cite{Wandzura:1977qf}, in the following calculation we ignore the contribution from the twist-3 TMD FFs $\tilde{D}_T$, $\tilde{H}_T^\perp$ and $\tilde{H}_{T}$.
Thus the structure functions $F_{UUT}^{\sin\phi_{S_h}}$ may be expressed as
\begin{align}
F_{UUT}^{\sin\phi_{S_h}}\approx {2M\over Q}\mathcal{I}[-xhH_1+{\bm{k_T}\cdot\bm{p_T}\over 2MM_h}(-xf^\perp\,D^\perp_{1T}-xg^\perp G_{1T})].
\end{align}
$F_{UUT}^{\sin\phi_{S_h}}$ is contributed by $h\,H_{1}$, $f^\perp\,D^\perp_{1T}$ and $g^\perp G_{1T}$ terms.
Then the $P_{h\perp}$-dependent transverse SSA $A_{UUT}^{\sin\phi_{S_h}}(P_{h\perp})$ is given as follows
\begin{align}
A_{UUT}^{\sin\phi_{S_h}}(P_{h\perp}) &= \frac{\int dx \int dy \int dz \;\mathcal{C}_{\mathrm{UT}}\;F_{UUT}^{\sin\phi_{S_h}}}{\int dx \int dy \int dz \;\mathcal{C}_{\mathrm{UU}}\;F_{\mathrm{UUU}}}\;,\label{asy1}
\end{align}
where we have defined the kinematical factors
\begin{align}
\mathcal{C}_{\mathrm{UU}}&=\frac{1}{x y Q^2}\frac{y^2}{2(1-\varepsilon)}\Bigl( 1+ \frac{\gamma^2}{2x}
\Bigl),\\
\mathcal{C}_{\mathrm{UT}}&=\frac{1}{x y Q^2}\frac{y^2}{2(1-\varepsilon)}\Bigl( 1+ \frac{\gamma^2}{2x} \Bigr) \sqrt{2\varepsilon(1+\varepsilon)},
\end{align}
with the ratio $\varepsilon$ of longitudinal and transverse photon flux being $\varepsilon=\frac{1-y-\gamma^2y^2/4}{1-y+y^2/2+\gamma^2y^2/4}$.
The $x$-dependent and the $z$-dependent asymmetries can be defined in a similar way.

Finally, in our estimate we also consider the following kinematical constraints on the intrinsic transverse momenta of the initial quarks in our calculation~\cite{Boglione:2011wm}:
\begin{align}
\begin{cases}
\bm p_{T}^2\leq(2-x)(1-x)Q^2, ~~~\textrm{for}~~0< x< 1 
;\\
\bm p_{T}^2\leq \frac{x(1-x)} {(1-2x)^2}\, Q^2, ~~~~~~~~~~~~\textrm{for}~~x< 0.5.
\end{cases}\label{constraints}
 \end{align}
 The first constraint is obtained by requiring the energy of the parton to be less than the energy of the parent hadron; while the second constraint is given by the requirement that the parton should move in the forward direction with respect to the parent hadron~\cite{Boglione:2011wm}.
 There are two upper limits for $\bm k_T$ in the region $x<0.5$, the smaller one should be chosen at the same time.

\subsection{Asymmetries at JLab 12 GeV}

We estimate the SSA $A_{UUT}^{\sin\phi_{S_h}}(P_{h\perp})$ of the transversely polarized hadron production in SIDIS at JLab with a $12~\textrm{GeV}$ electron beam, the kinematical cuts adopted in our calculation are:
\begin{align}
&0.1 < x < 0.6,~~0.3< z < 0.7,~~Q^2 > 1 \textrm{GeV}^2,\nonumber\\
&W^2 > 4\, \textrm{GeV}^2,P_{h\perp} > 0.05 \,\textrm{GeV}.
\end{align}

\begin{figure}
  \includegraphics[width=0.32\columnwidth]{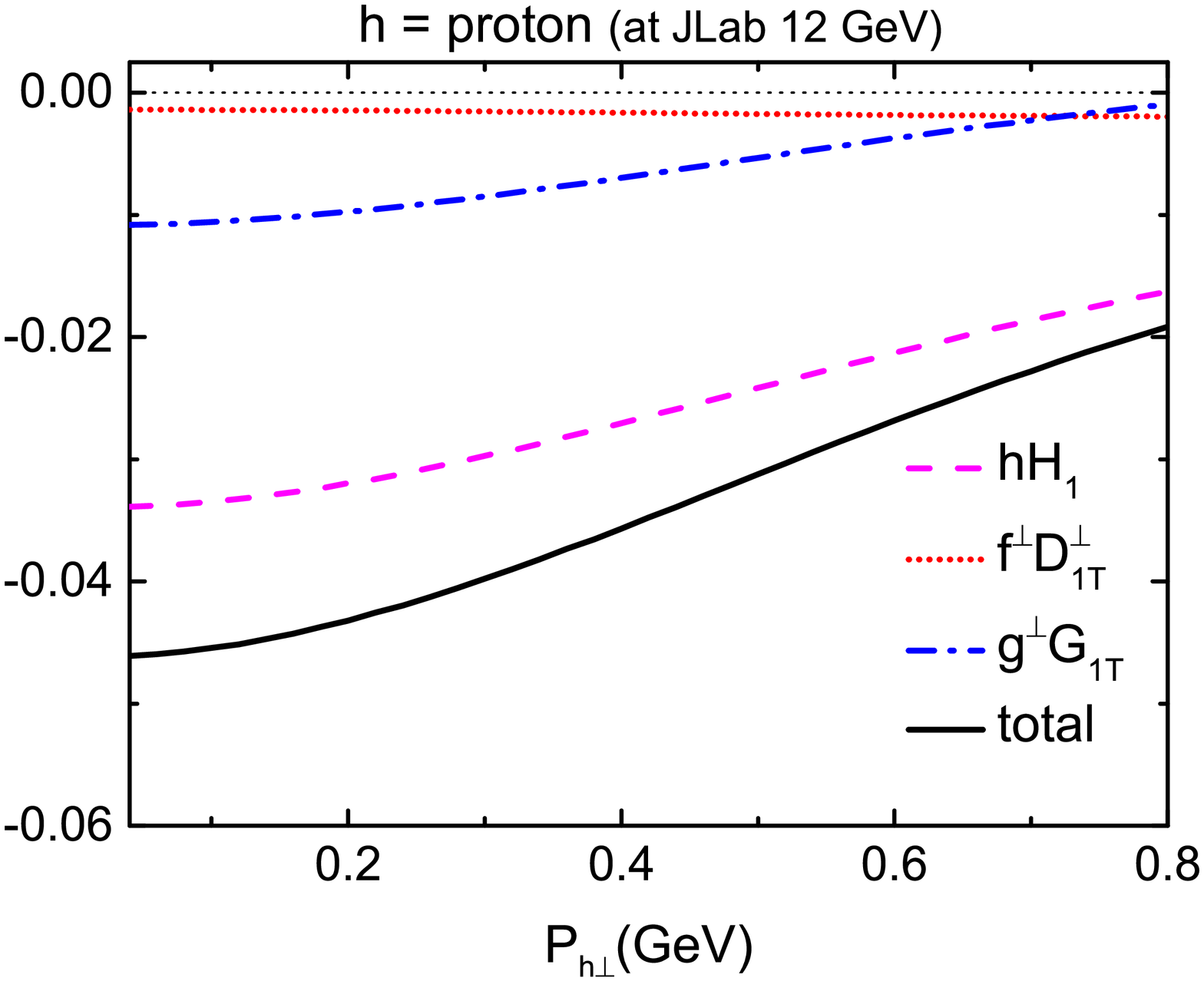}
  \includegraphics[width=0.32\columnwidth]{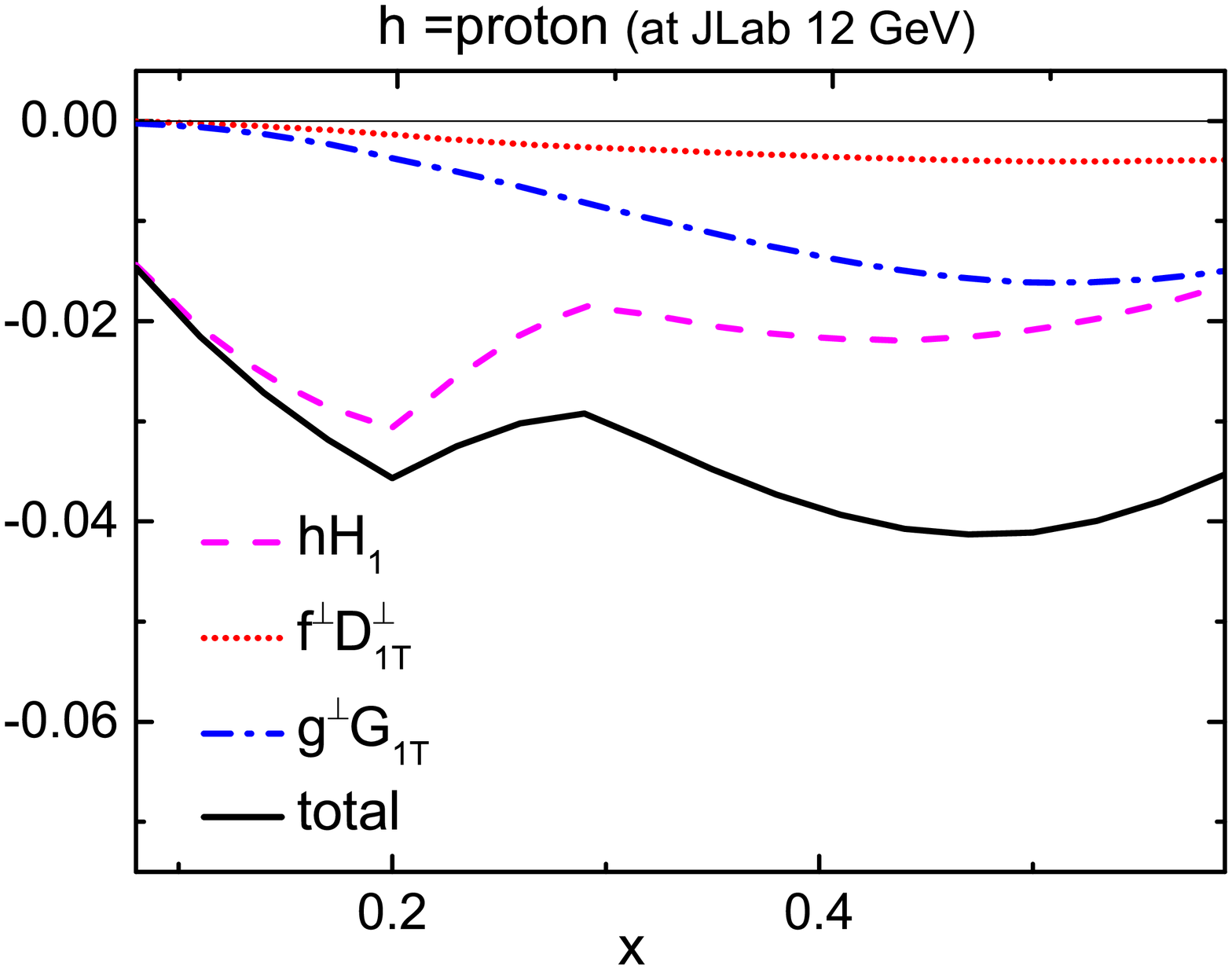}
  \includegraphics[width=0.32\columnwidth]{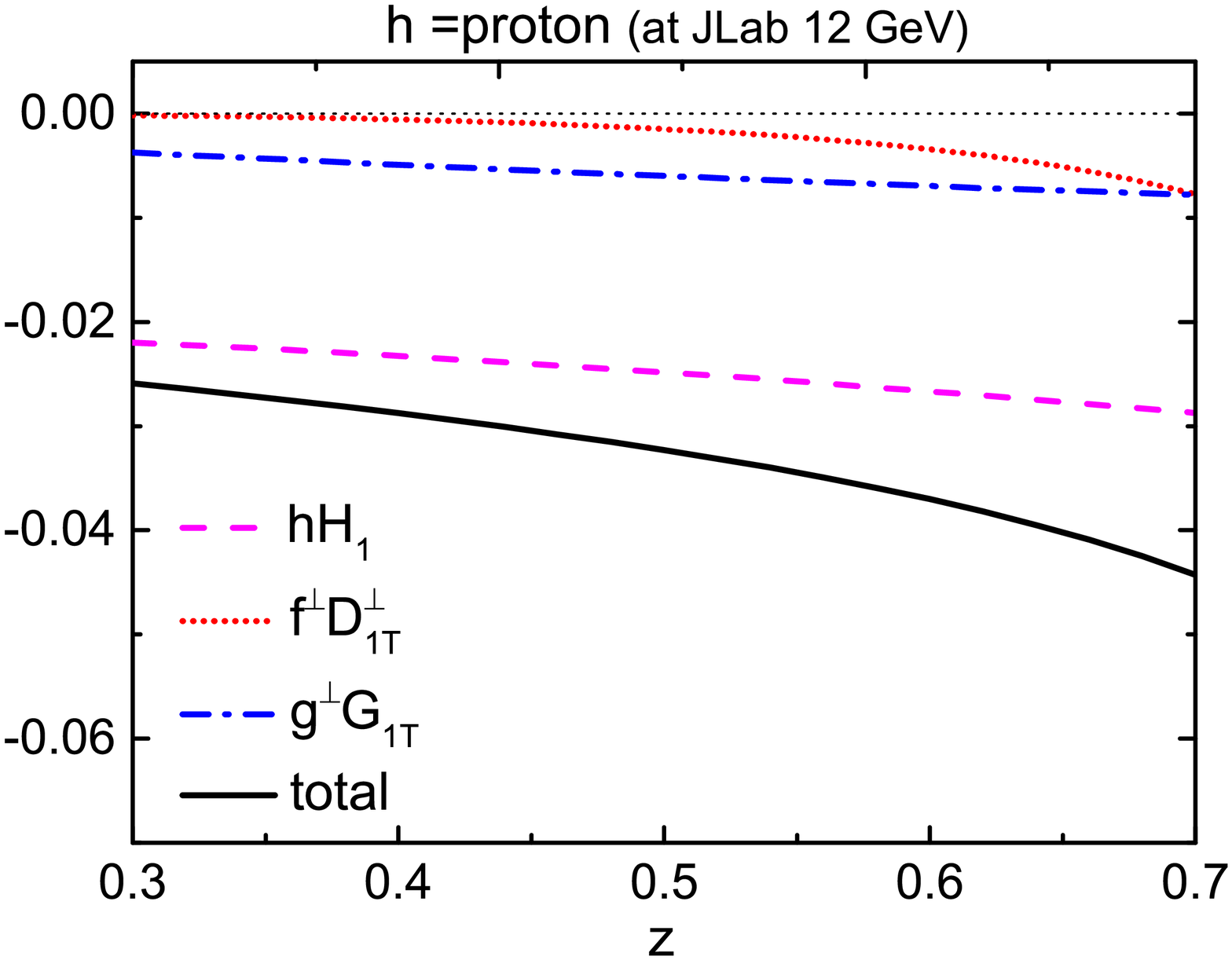}
  \caption {Predictions on the transverse SSA $A_{\mathrm{UUT}}^{\sin{\phi_{S_h}}}$ for proton production in SIDIS at JLab 12 GeV.
  The dashed, dotted and dash-dotted curves represent the asymmetries from the $hH_1$, $f^\perp D^\perp_{1T}$ and $g^\perp G_{1T}$ terms, respectively. The solid curves correspond to the total contribution.}
  \label{auut}
\end{figure}

In Fig.~\ref{auut}, we plot our prediction on $A_{\mathrm{UUT}}^{\sin\phi_{S_h}}$ at beam energy $12 \,\textrm{GeV}$ at JLab for the transversely polarized proton production as functions of $P_{h\perp}$, $x$ and $z$, respectively.
We find that the magnitude of the asymmetry $A_{\mathrm{UUT}}^{\sin\phi_{S_h}}$ for proton production is sizable and negative, the size is around $4\%$ at the kinematics of JLab.
In the $P_{h\perp}$-dependent and $z$-dependent asymmetries, the $h H_1$ term dominates over the $g^\perp G_{1T}$ and $f^\perp D_{1T}^\perp$ terms.
The contribution from the $f^\perp D_{1T}^\perp$ term is much smaller and can be almost neglected.
This is due to the very small size of $D_{1T}^\perp$ of the proton in the moderate $z$ region.
Thus the $\sin\phi(S_h)$ asymmetry in transversely polarize proton production may provide an opportunity to probe the unknown TMD distribution $h(x,\bm p_T^2)$ and the fragmentation function $H_1(z,\bm k_T^2)$.

\begin{figure}
  \includegraphics[width=0.32\columnwidth]{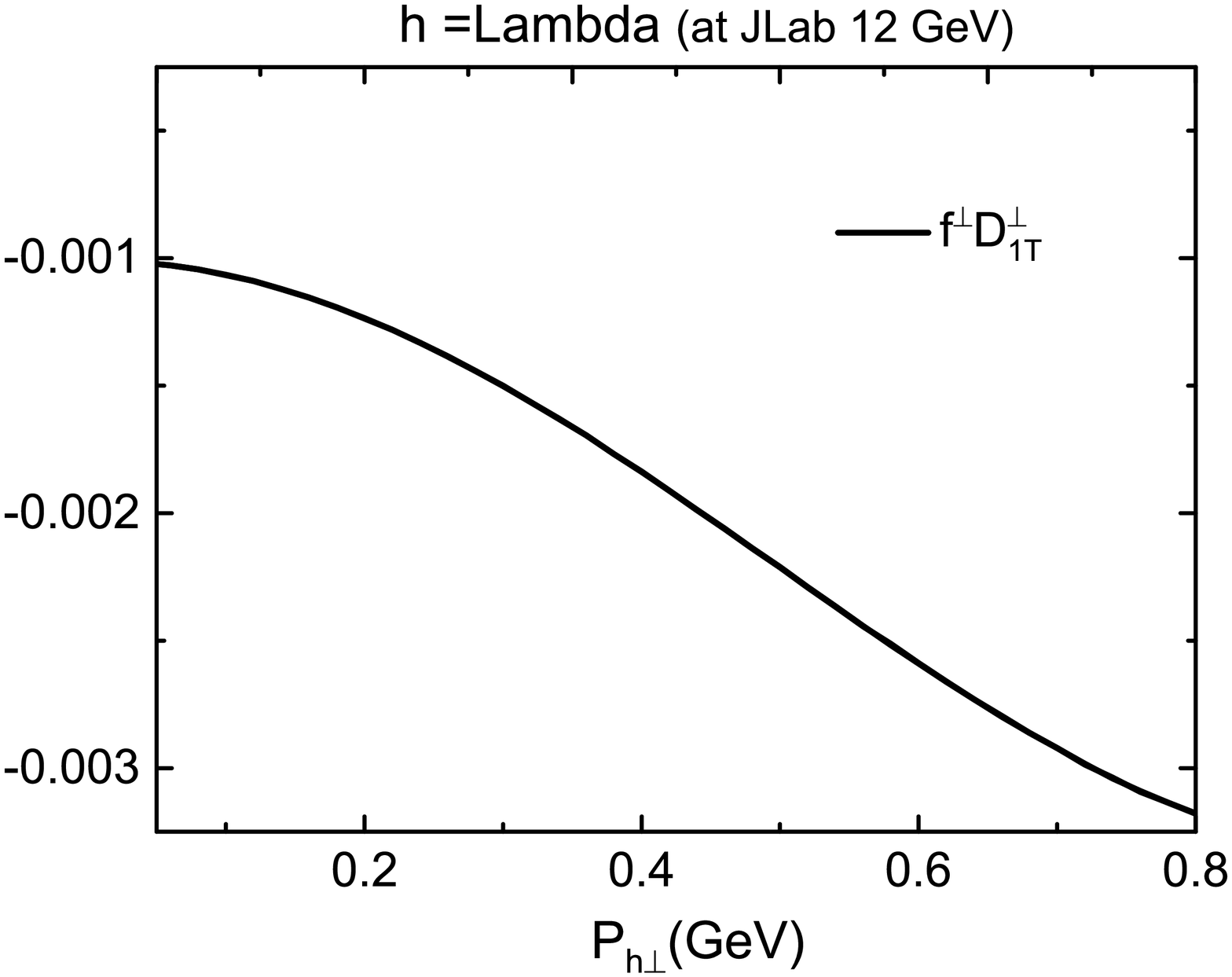}
  \includegraphics[width=0.32\columnwidth]{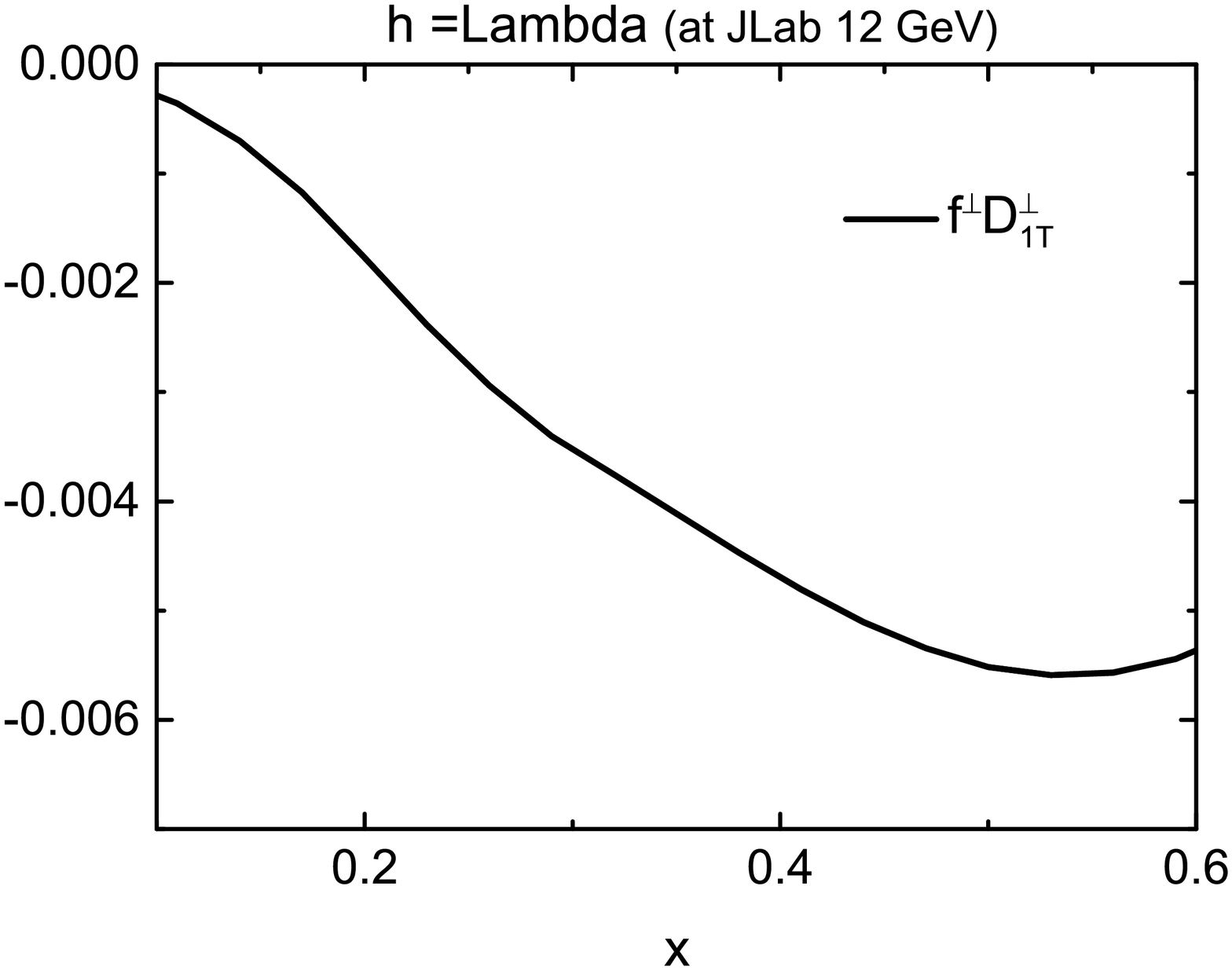}
  \includegraphics[width=0.32\columnwidth]{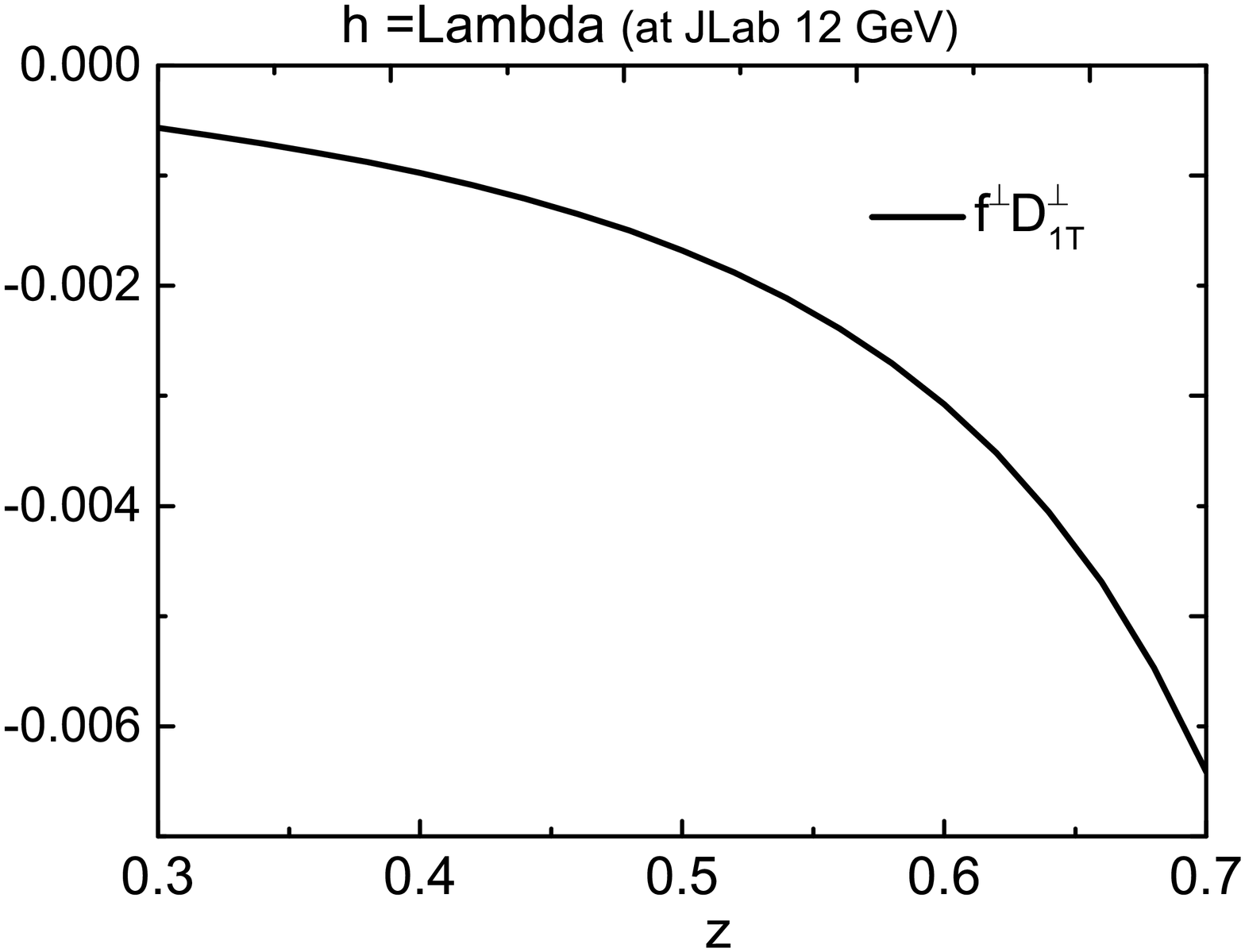}
  \caption {Predictions on the transverse SSA $A_{\mathrm{UUT}}^{\sin{\phi_{S_h}}}$ for Lambda production in SIDIS at JLab.
  The solid curves correspond to the total asymmetry (it only receives contribution from the $f^\perp D^\perp_{1T}$ term).}
 \label{aul}
\end{figure}

We also predict the asymmetry $A_{\mathrm{UUT}}^{\sin\phi_{S_h}}$ at JLab $12 \textrm{GeV}$ for the transversely polarized $\Lambda$ hyperon production as functions of $P_{h\perp}$, $x$ and $z$, respectively.
In our model for the fragmentation functions, $H_1$ and $G_{1T}$ for the $u$ and $d$ quarks are vanishing.
The T-even fragmentation functions only receives the contribution from the strange quark.
Thus, the $hH_1$ and $g^\perp G_{1T}$ terms are zero as in our spectator model $h$ and $g^\perp$ of the strange quark also vanish.
Only the term $f^\perp D^\perp_{1T}$ survives in the asymmetry $A_{\mathrm{UUT}}^{\sin\phi_{S_h}}$.
The numerical results of $A_{\mathrm{UUT}}^{\sin\phi_{S_h}}$ vs $x$, $z$, and $P_{h\perp}$ are plotted in the Fig.~\ref{aul}.
We can find that the asymmetry $A_{\mathrm{UUT}}^{\sin\phi_{S_h}}$ for $\Lambda$ hyperon production is smaller than 1$\%$, which is smaller compared to that for the proton production.
Nevertheless, the $\sin\phi_{S_h}$ asymmetry in transversely lambda production may provide a clean way to access the $f^\perp D_{1T}^\perp$ as there are no competing terms in this process.

\subsection{Asymmetries at COMPASS}

\begin{figure}
  \includegraphics[width=0.32\columnwidth]{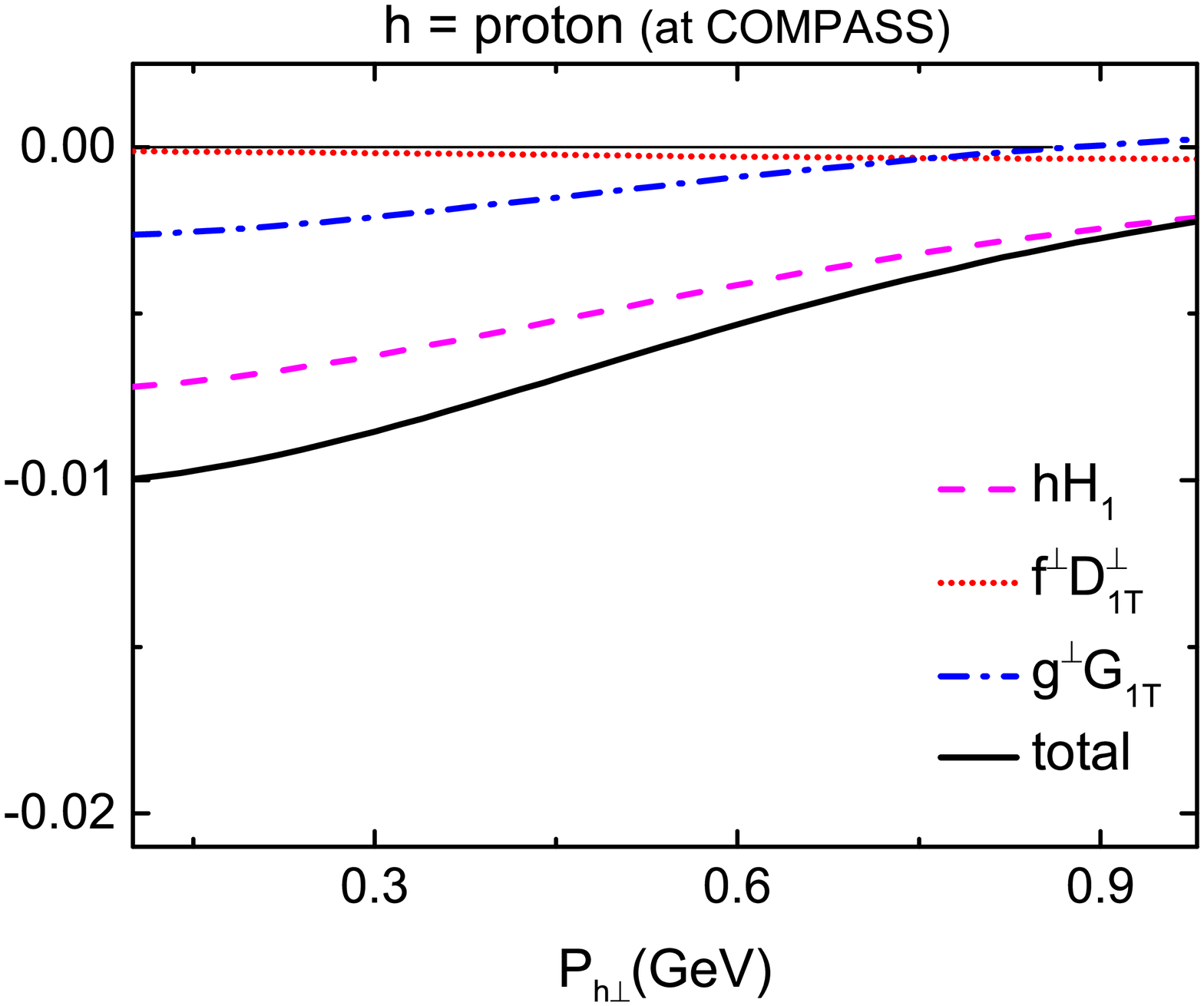}
  \includegraphics[width=0.32\columnwidth]{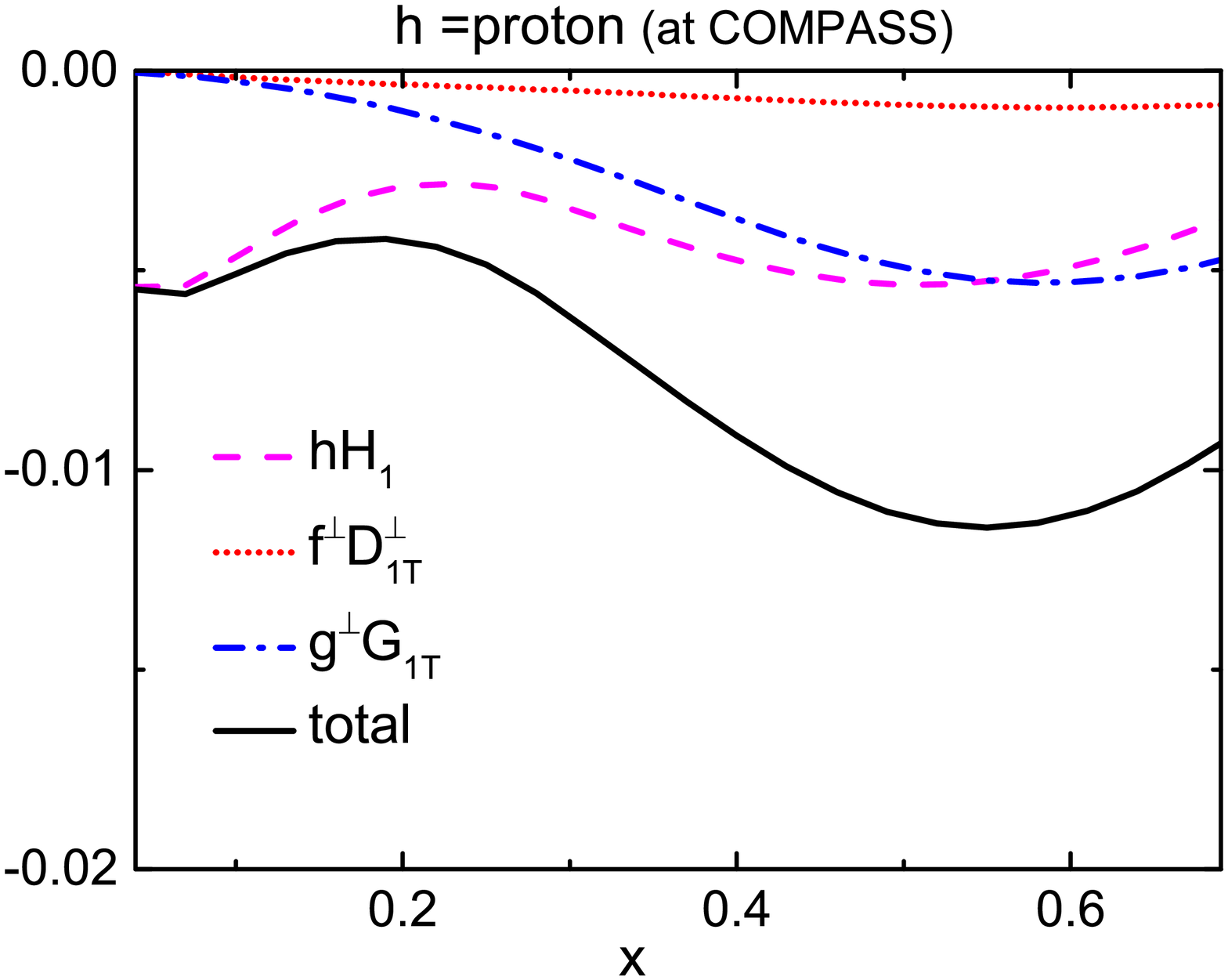}
  \includegraphics[width=0.32\columnwidth]{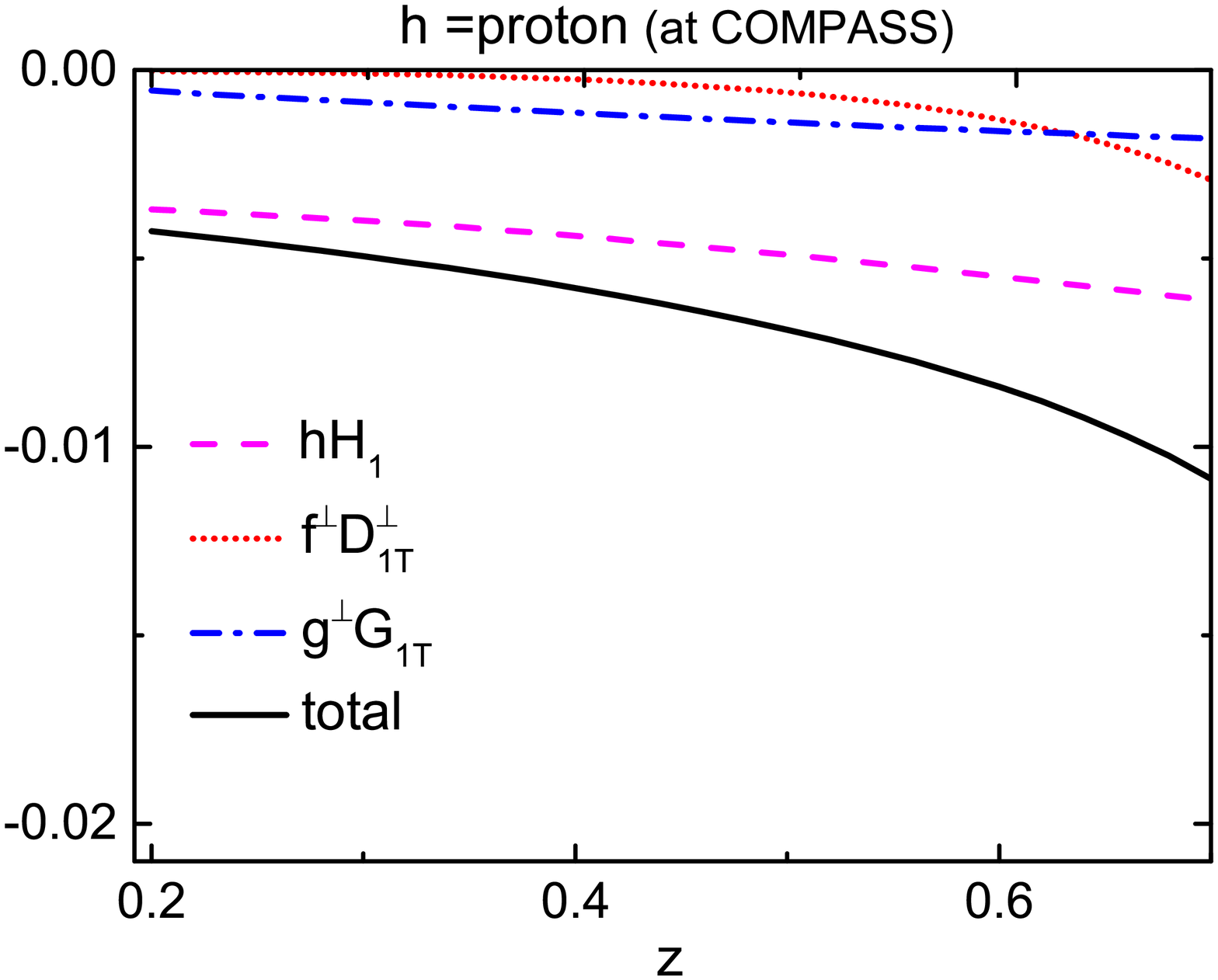}
  \caption {Predictions on the transverse SSA $A_{\mathrm{UUT}}^{\sin{\phi_{S_h}}}$ for the proton production in SIDIS at COMPASS.
  The dashed, dotted and dash-dotted curves represent the asymmetries from the $hH_1$, $f^\perp D^\perp_{1T}$ and $g^\perp G_{1T}$ terms, respectively.
The solid curves correspond to the total contribution.}
  \label{auutc}
\end{figure}

\begin{figure}
  \includegraphics[width=0.32\columnwidth]{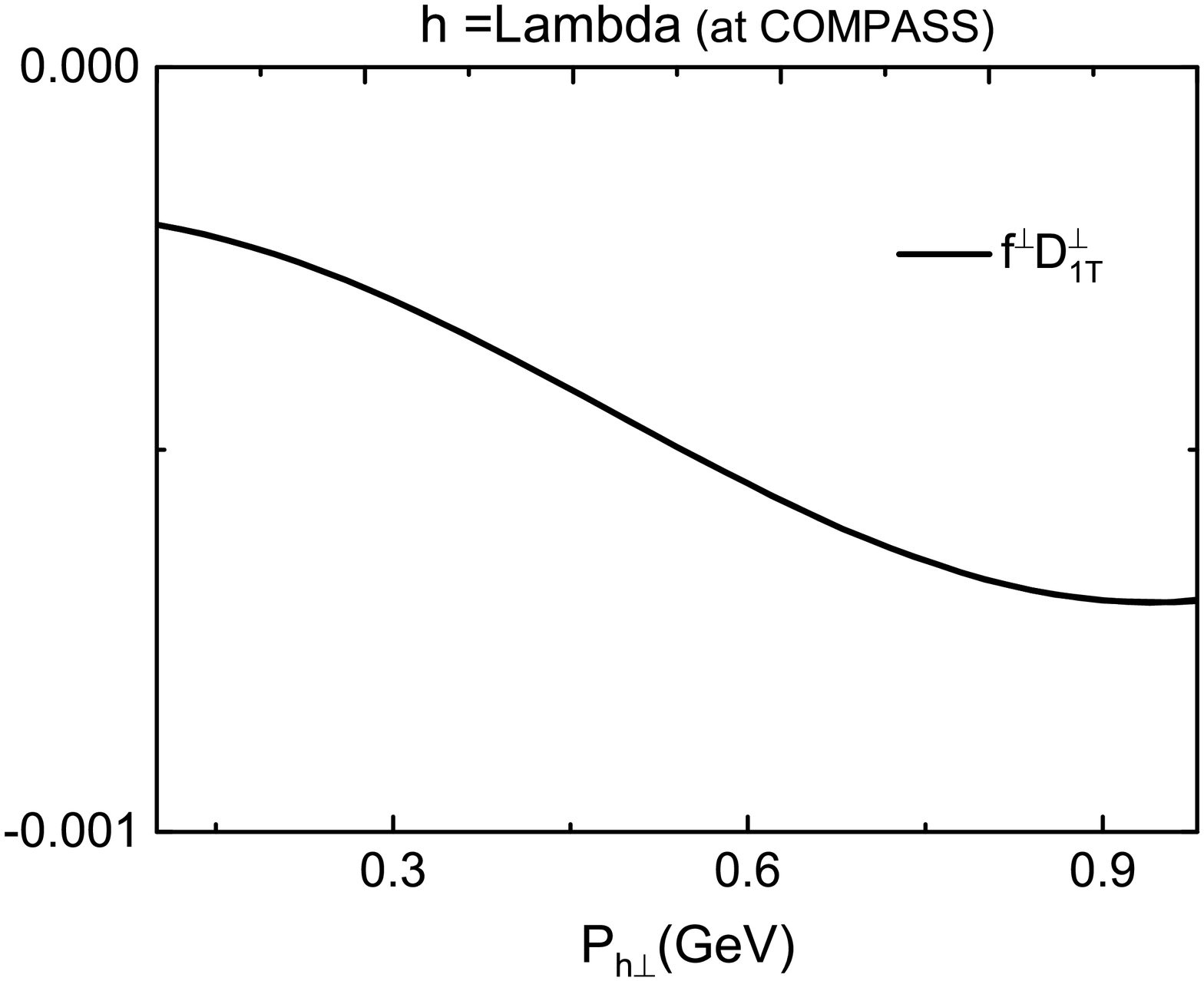}
  \includegraphics[width=0.32\columnwidth]{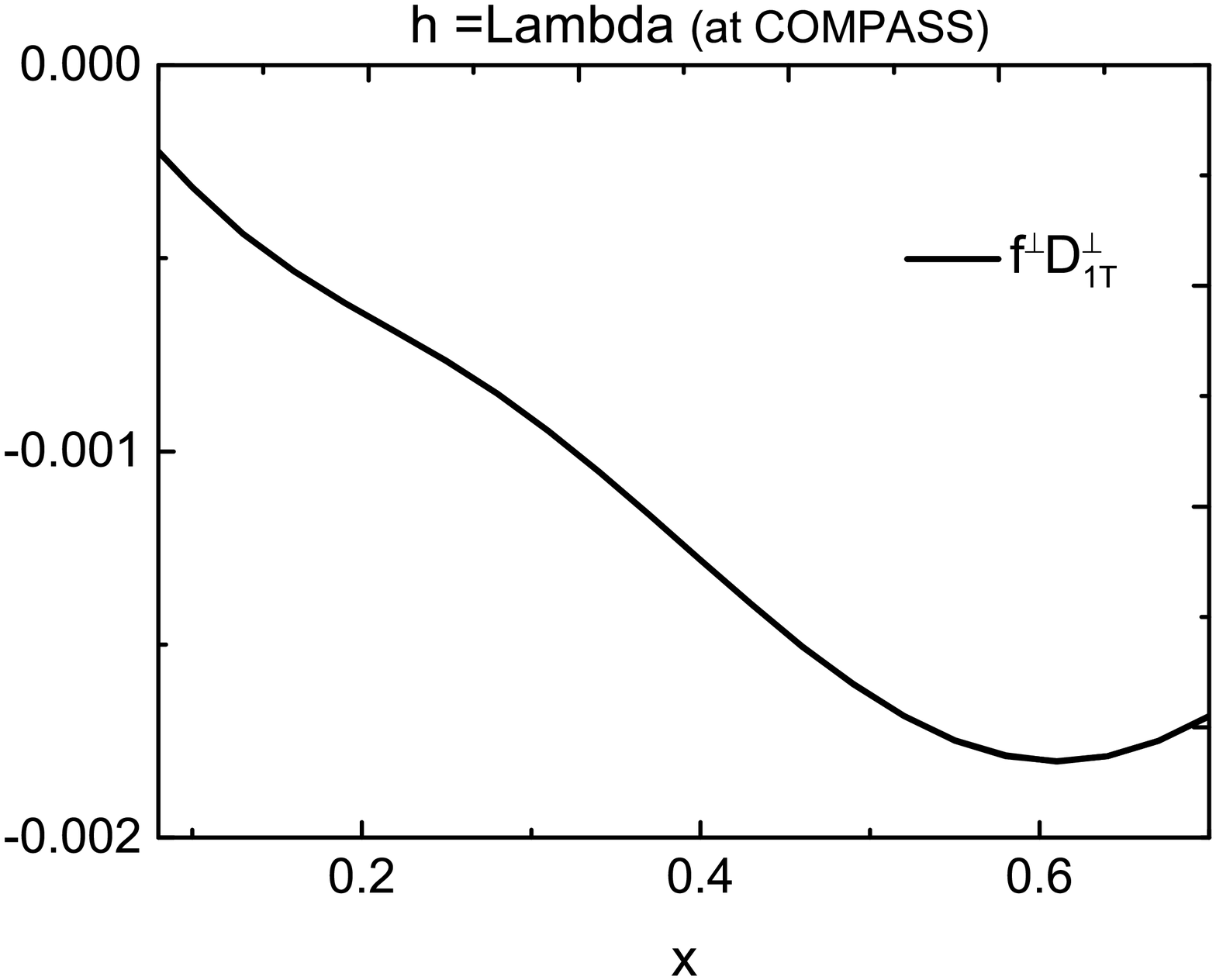}
  \includegraphics[width=0.32\columnwidth]{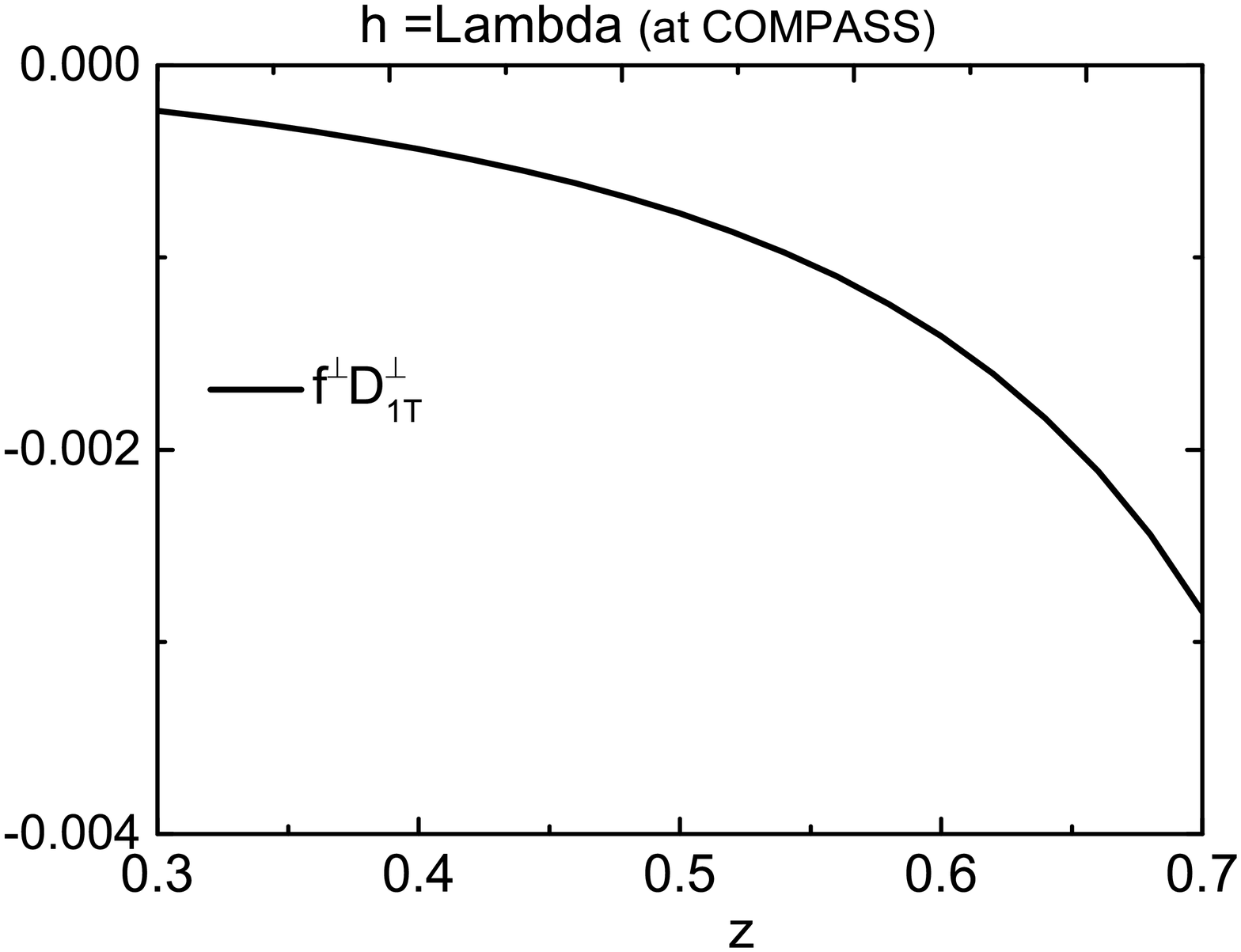}
  \caption {Predictions on the transverse SSA $A_{\mathrm{UUT}}^{\sin{\phi_{S_h}}}$ for Lambda production in SIDIS at COMPASS.
  The solid curves correspond to the total asymmetry (it only receives contribution from $f^\perp D^\perp_{1T}$ term in our model).}
 \label{aulc}
\end{figure}

To study the energy dependence of the asymmetries, we also estimate the transverse asymmetries for proton and Lambda at COMPASS with a muon beam of $160~\text{GeV}$
as a further comparison.
In this calculation, we adopt the following kinematical cults~\cite{Alekseev:2010rw}:
\begin{align}
&0.004<x<0.7,~~ 0.1<y<0.9,~~ z > 0.2,\nonumber\\
&P_{h\perp}>0.1\,\textrm{GeV},~~Q^2>1\, \textrm{GeV}^2,\nonumber\\
&W>5\,\textrm{GeV}, ~~ E_h > 1.5\, \textrm{GeV}.
\end{align}
The numerical results of the asymmetries $A_{\mathrm{UUT}}^{\sin{\phi_{S_h}}}$ for proton and Lambda hyperon are shown in Figs.~\ref{auutc} and~\ref{aulc}, respectively.

Our prediction for the $\sin{\phi_{S_h}}$ of the  proton production show that the $hH_1$ term dominates over the $g^\perp G_{1T}$ and $f^\perp D_{1T}^\perp$ terms at COMPASS,
similar to the case at JLab.
We also find that the size of the $hH_1$ contribution is around $1\%$ for the proton production at the kinematics of COMPASS, and it is clearly smaller than that of JLab.
This is because that $Q$ at COMPASS is larger than that at JLab, and the twist-3 effect is suppressed by a factor of $1/Q$.
The asymmetries for Lambda hyperon is also smaller than that at JLab.

\section{conclusion}

In this work, we studied the role of the twist-3 TMD distribution and twist-2 fragmentation functions in the $sin(\phi_{S_h})$ azimuthal asymmetry in SIDIS process,
in which the final state hadron is a transversely polarized proton or lambda hyperon.
We preformed the model calculation of the twist-3 TMD distributions $h$ and $f^\perp$ as well as the twist-2 fragmentation functions $H_1$, $G_{1T}$ and $D^\perp_{1T}$
using the spectator diquark model.
In calculating the TMD distributions, we considered the both scalar and axial-vector diquarks~\cite{Jakob:1997wg,Bacchetta:2008af}, respectively.
Using the model results on the TMD distribution and fragmentation functions, we predicted the SSA $A_{UUT}^{\sin\phi_{S_h}}$ for the transversely polarized
proton and Lambda production in SIDIS at the kinematics of JLab 12 GeV and at COMPASS.
We find that the estimated the asymmetry $A_{UUT}^{\sin\phi_{S_h}}$ for the proton is sizable.
Specifically, the magnitude is around 4 percent at JLab 12 GeV, while it is about 1 percent at COMPASS.
The asymmetries $A_{UUT}^{\sin\phi_{S_h}}$ for the lambda hyperon is much smaller.

For more discussion, we also compared the contributions to the asymmetry $A_{UUT}^{\sin\phi_{S_h}}$ from different origins: the $hH_1$, $g^\perp G_{1T}$ and $f^\perp D^\perp_{1T}$ terms. The comparison showed that the $h H_1$ term dominates in the transversely polarized proton production, while the $f^\perp D^\perp_{1T}$ term is almost negligible in this case.
In contrast, in the $\sin\phi_{S_h}$ asymmetry of the transversely polarized lambda production, only the $f^\perp D^\perp_{1T}$ term survives in our model. Although this asymmetry is small, it might be still measurable at the kinematics of JLab 12 GeV.
Our study shows that the measurement of the $\sin\phi_{S_h}$ in the proton production at JLab and COMPASS is feasible to probe the unknown TMD distribution $h$ and fragmentation function $H_1$.
We note that independent information of $H_1$ may be also accessible in the electron-positron annihilation process $e^+ e^- \to h_1^\uparrow h_2^\uparrow X$, which can be combined with the data in SIDIS to perform the phenomenological analysis.

\section{Acknowledgements}
This work is partially supported by the National Natural Science Foundation of China (NSFC) grants No.~11575043 and No.~11605297, and by the Fundamental Research Funds for the Central Universities of China.
Y.~Yang is supported by the Scientific Research Foundation of Graduate
School of Southeast University (Grant No. YBJJ1770) and by the Postgraduate Research \& Practice Innovation Program of Jiangsu Province (Grant No. KYCX17\_0043).
W.~Mao is also supported by the High-level Talents Research and Startup Foundation Projects for Doctors of Zhoukou Normal University (ZKNUC2016014),
by the Fundamental Research Funds for the Central Universities of China.

\end{document}